\documentclass[fleqn,usenatbib]{mnras}

\usepackage{newtxtext,newtxmath}
\usepackage[T1]{fontenc}
    
\DeclareRobustCommand{\VAN}[3]{#2}
\let\VANthebibliography\thebibliography
\def\thebibliography{\DeclareRobustCommand{\VAN}[3]{##3}\VANthebibliography}

\usepackage{graphicx}	
\usepackage{amsmath}	
\usepackage{cleveref}

\title[EM counterpart of spinning BNS]{The imprint of individual neutron star spins on ejecta, $r$-process nucleosynthesis, and kilonovae in binary neutron star mergers}

\author[Karakaş, Matur, Radice, Haas, Ruffert]{
Beyhan Karakaş,$^{1}$\thanks{E-mail: \href{mailto:beyhannkarakas@gmail.com}{beyhannkarakas@gmail.com}}
Rahime Matur,$^{2}$\thanks{E-mail: \href{mailto:r.matur@soton.ac.uk}{r.matur@soton.ac.uk}}
David Radice,$^{3,4,5}$
Roland Haas,$^{6,7,8}$
Maximilian Ruffert$^{9}$
\\
$^{1}$beyhannkarakas@gmail.com\\
$^{2}$Mathematical Sciences and STAG Research Centre, University of Southampton, Southampton SO17 1BJ, UK\\
$^{3}$Institute for Gravitation and the Cosmos, The Pennsylvania State University, University Park, PA 16802, USA\\
$^{4}$Department of Physics, The Pennsylvania State University, University Park, PA 16802, USA\\
$^{5}$Department of Astronomy \& Astrophysics, The Pennsylvania State University, University Park, PA 16802, USA\\
$^{6}$Department of Physics and Astronomy, University of British Columbia, Vancouver, British Columbia, Canada\\
$^{7}$National Center for Supercomputing applications, University of Illinois, Urbana, Illinois, USA\\
$^{8}$Department of Physics, University of Illinois, Urbana, Illinois, USA\\
$^{9}$School of Mathematics and Maxwell Institute, University of Edinburgh, Edinburgh EH9 3FD, UK
}

\date{Accepted XXX. Received YYY; in original form ZZZ}
\pubyear{\the\year{}}

\begin{document}
\label{firstpage}
\pagerange{\pageref{firstpage}--\pageref{lastpage}}
    \maketitle

\begin{abstract}
To leading order, the gravitational-wave (GW) signal from binary neutron stars depends on the individual spins, $\chi_1$ and $\chi_2$, only through the effective spin parameter $\chi_{\rm eff}$. We present the first systematic investigation of individual-spin effects on ejecta, $r$-process nucleosynthesis, and kilonova emission, including comparisons at fixed total mass, mass ratio and $\chi_{\rm eff}$. We use numerical relativity ejecta from three total mass regimes with the finite-temperature, composition-dependent SFHo equation of state and neutrino emission and absorption. For $M_{\rm tot}=2.55\,M_\odot$ at fixed $\chi_{\rm eff}=0$, individual spins change the dynamical ejecta mass by a factor of ${\sim}45$, while the absolute $A\geq140$ yield spans more than two orders of magnitude and the lanthanide to light $r$-process mass ratio increases from ${\sim}2$ to ${\sim}70$. Prompt-collapse $4.10\,M_\odot$ models show heavy-element yield differences exceeding four orders of magnitude at $\chi_{\rm eff}=0$. The kilonova retains the individual-spin imprint, with peak brightness differences reaching ${\sim}0.9$ mag. At $40\,{\rm Mpc}$, all three fixed $\chi_{\rm eff}=0$ pairs remain above adopted depths at common epochs for all viewing angles, with same-epoch colour differences reaching ${\sim}1.5$ mag. The colour imprint persists when the simulation-derived secular ejecta are replaced by the same parametric disc outflow, indicating that disc mass differences are not the primary driver and that dynamical ejecta make an important contribution. Neutrino absorption systematically brightens the kilonova and shifts peak-associated colours blueward. These EM signatures can break the degeneracy between individual spins in the GW signal.
\end{abstract}

\begin{keywords}
stars: neutron -- stars: rotation -- neutrinos -- transients: neutron star mergers -- nuclear reactions, nucleosynthesis, abundances -- hydrodynamics
\end{keywords}

\section{Introduction}

The gravitational waves (GWs) from the binary neutron star (BNS) merger \texttt{GW170817}~\citep{gw170817_ligo} were followed by an electromagnetic (EM) counterpart detected from $\gamma$-rays to radio frequencies, making it the first and, to date, the only confirmed multi-messenger observation of a compact binary merger~\citep{gw170817_multimessenger}. A short gamma-ray burst (sGRB)~\citep{sgrb_1992}, \texttt{GRB170817A}, was detected $1.74^{+0.05}_{-0.05}\, \mathrm{s}$ after the merger by the \emph{Fermi Gamma-ray Burst Monitor}~\citep{grb170818a_fermi, grb170817a_ligo_fermi_integral} and independently by \emph{INTEGRAL}~\citep{grb170817a_integral}. The source was localized to the galaxy NGC 4993 at a distance of ${\sim}40\, \mathrm{Mpc}$~\citep{sss17a_optical, grb170817a_ngc4993}. 

\texttt{GW170817} was the closest GW event observed, which enabled the detection of its EM counterpart. However, despite this proximity, \texttt{GRB170817A} was an off-axis sGRB~\citep{grb170817a_xray_troja, grb170817a_offaxis, at2017gfo_radio, at2017gfo_uv_xray}, with a fluence at least a factor of $100$ lower than that of typical observed sGRBs~\citep{grb170817a_ligo_fermi_integral}. Its afterglow was detected in X-rays and radio at $9$ and $16$ days after the merger, respectively, and peaked at ${\sim}150$ days~\citep{grb170817a_xray_troja, at2017fgo_xray, at2017gfo_radio, at2017gfo_uv_xray, at2017gfo_superluminal}. Following the detection of~\texttt{GRB170817A}, the thermal transient \texttt{AT2017gfo} was first identified in the optical at $10.98$ hours after the merger~\citep{sss17a_optical, at2017gfo_optik_dlt, unprecedented_2017, optical_2017}, with subsequent ultraviolet and near-infrared observations~\citep{at2017gfo_uv_xray, at2017gfo_nir}.
\texttt{AT2017gfo} was consistent with thermal emission powered by the radioactive decay of rapid neutron capture process ($r$-process) elements, i.e. a kilonova (see~\cite{metzger_review} for a review) for which the first detailed light curves and spectroscopic observations were obtained, including the identification of strontium~\citep{strontium}.

The joint observation of \texttt{GW170817} and \texttt{GRB170817A} provided the first observational confirmation that BNS mergers can act as progenitors of at least some sGRBs~\citep{ grb170817a_ligo_fermi_integral, grb170817a_integral}. The associated kilonova, \texttt{AT2017gfo}, revealed that the EM emission requires at least two distinct ejecta components with different compositions, masses and velocities~\citep{at2017gfo_nir, rapid_reddening_2017, optical_2017, rprocess_spectra_2017, kasen_2017, early_spectra_2017}. The early time emission was dominated by a blue component, characterised by relatively fast-moving and low opacity, consistent with lanthanide-poor ejecta (corresponding to the production of light $r$-process elements with $A \lesssim 140$) ~\citep{grb170817a_integral, at2017gfo_optik_dlt, at2017gfo_nir, grb170817a_ligo_fermi_integral}. This blue component is inferred to consist of a relatively low ejecta mass $M_{\mathrm{ej}} {\sim}3\times10^{-3} - 10^{-2}\, M_{\odot}$ moving at ${\sim}0.2-0.3\, c$, with low opacity ($\kappa \lesssim 1\, \mathrm{cm^{2}\, g^{-1}}$) indicating a lanthanide-poor composition (with a lanthanide mass fraction $X_{\mathrm{lan}} {\sim}10^{-4.5} - 10^{-4}$)~\citep{grb170817a_integral, at2017gfo_optik_dlt, at2017gfo_nir, grb170817a_ligo_fermi_integral}. At later times, approximately three days after the merger, the emission became dominated by a red component associated with more massive, slower moving ejecta and higher opacity, consistent with a lanthanide-rich composition (corresponding to the production of heavy $r$-process elements with $A \gtrsim 140$)~\citep{at2017gfo_optik_dlt, at2017gfo_nir}. This red component is inferred to have ejecta masses of $M_{\mathrm{ej}} {\sim}10^{-2} - 10^{-1}\, M_{\odot}$ moving at ${\sim}0.05 - 0.1\, c$ with high opacity ($\kappa {\sim}10\, \mathrm{cm^{2}\, g^{-1}}$) consistent with a lanthanide-rich composition ($X_{\mathrm{lan}} {\sim}10^{-2} - 10^{-1}$)~\citep{grb170817a_integral, at2017gfo_optik_dlt, at2017gfo_nir, kasen_2017, Villar_2017, Perego_2017, Radice_joint2018, Kawaguchi_2018}.

Prior to these observations, several theoretical and numerical studies had predicted that BNS mergers could act as progenitors of sGRBs~\citep{eichler_nucleo_1989, piran_grb_1992, ruffert_1994_gammaray, Ruffert_coalescing_1995, Ruffert_coalescing_1996, Ruffert_coalescing_1997, ruffert_1998, ruffert_grb_1998, ruffert_grb_1999, Ruffert_coalescing_2001, metzger_sgrb, nagakura_sgrb_2014}, play a significant role in forming $r$-process elements~\citep{rprocess_evans_1988, rprocess_meyer_1989, freiburghaus_rprocess_1999, radice_massejection_2016}, and produce EM counterparts~\citep{li_pacyznski_1998, metzger_2010_em, korobkin_rprocess_2012, metzger_em_2012, Hotokezeka_kilonova_2013, rosswog_2013_rprocess_em, metzger_red_blue_2014, metzger_neutronpowered_2015, kasen_discwind_2015, tanaka_kilonova_2016, hotokezaka_em_2016, dietrich_dynejectafit_2016}.

Following \texttt{GW170817}, numerical relativity simulations were used to interpret the inferred ejecta properties and their dependence on weak interactions, equations of state (EoSs) and binary parameters~\citep{Radice_viscousdyn_2017, Radice_longlived_2018, radice_gw170817eos_2018, radice_massejection_2018, radice_viscousdyn_2018, radice_viscousdyn_2020, 
Kawaguchi_2018, kawaguchi_diversity_2020, kawaguchi_lowmass_2021, Kawaguchi_shortlived_2023, Fujibayashi_lowmass_2020} (see for reviews on mass ejection~\cite{shibata_review_2019}, on dynamics~\cite{radice_review_2020}, modeling gravitational waves~\cite{dietrich_review_2021}, neutrino transport~\cite{foucart_review_2023}, and turbulence modelling~\cite{radice_turbulence_review_2024} in BNS mergers). These studies established that BNS mergers produce multiple ejecta components through distinct mechanisms, including dynamical ejecta arising from shock heating, tidal interactions, additional viscous-dynamical ejecta in unequal mass binaries~\citep{Radice_viscousdyn_2017, radice_viscousdyn_2018, radice_viscousdyn_2020}, and secular outflows driven by neutrino irradiation, viscosity and spiral-wind~\citep{Siegel_3d_grmhd_2017, Fernandez_longterm_2018, Miller_gw170817_2018, Siegel_gw170817_2019,  spiralwind_nedora, Kiuchi_second2023, Radice_secular_24, matur_bhns_spin_2025}. The masses, velocities, compositions and angular distributions of these components were shown to account for the observed blue and/or red components of \texttt{AT2017gfo}. These studies further showed that the lifetime of the BNS remnant plays a key role in shaping the EM signal, with prompt or short-lived NS remnants that collapse to a BH leading to fainter and shorter-lived optical emission, while long-lived NS remnants can sustain polar, lanthanide-poor outflows consistent with the observed blue component~\citep{Radice_longlived_2018, radice_massejection_2018, Kawaguchi_2018, Kawaguchi_shortlived_2023}. At the same time, simulations highlighted degeneracies between microphysical inputs and binary parameters, indicating that not all properties of the system can be uniquely constrained from EM observations alone~\citep{radice_gw170817eos_2018, radice_massejection_2018, Kawaguchi_linking_2025}.

Previous studies of spin effects in BNS mergers were reviewed in our previous paper~\citep{effofspin}. Overall, studies including NS spin remain very limited compared to the large number of studies assuming irrotational binaries. Among these studies, only a small number explored the consequences of spin for $r$-process nucleosynthesis or kilonova observables~\citep{dietrich_spin_2017, East_spin_2019, papenfort_extreme_2022, rosswog_spin_2024, dietrich_spin26, Allen_jay_spin26}. \citet{dietrich_spin_2017, East_spin_2019, rosswog_spin_2024} employed piecewise-polytropic EoSs, while \citet{papenfort_extreme_2022} used finite-temperature, composition-dependent EoSs, but neglected neutrino absorption and restricted the analysis to single-spin aligned models. More recent studies including magnetic fields and neutrino transport considered equal mass binaries with aligned spins at $\chi = 0.1$~\citep{dietrich_spin26}, while the same masses were also studied at higher spins, $|\chi|=0.43$, for aligned and anti-aligned configurations~\citep{Allen_jay_spin26}. The former is restricted to a single low spin model, while the latter does not present kilonova calculations. In these studies, varying the individual spins, $\chi_1$ and $\chi_2$, also changes $\chi_\mathrm{eff}$. The effect of varying $\chi_1$ and $\chi_2$ separately at fixed $\chi_\mathrm{eff}$ on the relative contributions of different ejecta channels and the resulting $r$-process nucleosynthesis and kilonova signatures therefore remains unexplored.

In this work, we present the first systematic investigation of how the individual spin configuration affects the ejecta, $r$-process nucleosynthesis, and kilonova emission in BNS mergers. In particular, we test whether binaries with the same effective spin, $\chi_\mathrm{eff} = (M_1 \chi_1 + M_2 \chi_2)/(M_1 + M_2)$, but different individual spin configurations produce equivalent EM counterparts. We consider irrotational, aligned, anti-aligned, single-spin aligned, single-spin anti-aligned and mixed configurations across three total mass regimes. All models have $q=1$, except for $q=2.05$ configurations at $M_\mathrm{tot}=3.05\,M_\odot$. Our models employ a finite-temperature, composition-dependent EOS with neutrino emission and absorption, and we analyse the tidal, shock-heated, high-velocity, fast and secular ejecta to determine how the spin dependent differences in these ejecta channels affect the nucleosynthesis and kilonova signatures. For several spin configurations, we perform otherwise identical simulations with and without neutrino absorption to isolate the effect of neutrino absorption from that of spin. We show that binaries with the same $\chi_\mathrm{eff}$, but different individual spins can produce different ejecta, $r$-process nucleosynthesis and kilonova emission. The EM counterpart can therefore break the individual spin degeneracy in the GW signal and could be used to constrain the formation channels of these binaries.

The paper is organized as follows. In~\cref{setup} we describe the numerical setup and analysis methods.~\Cref{sec:ejecta} examines the impact of spin on the properties of different ejecta components.~\Cref{sec:nucleosynthesis} presents the $r$-process yields.~\Cref{sec:kilonovae} connects these spin-dependent ejecta properties to EM observables, and~\cref{conclusion} summarizes the main conclusions.

\section{Analysis and setup}\label{setup}

In this section, we analyze the outflow data from the simulations presented in~\citep{effofspin}. Details of the numerical setup are given therein. All models are evolved with the finite-temperature composition-dependent SFHo EoS~\citep{steiner_core-collapse_2013} from \texttt{StellarCollapse}~\citep{stellarcollapse_2010}.
Here we summarize only the elements relevant for the ejecta, nucleosynthesis and kilonova modeling.

The general relativistic hydrodynamic evolution is performed with \texttt{WhiskyTHC}~\citep{radice_thc_2012, radice_highorder_2014, radice_beyond_2014, radice_thc_2015} and spacetime evolution is carried out with \texttt{CTGamma}~\citep{pollney_high_2011}. Weak interactions are modeled with the M0+Leakage scheme as described in~\cite{effofspin}. For $M_\mathrm{tot} = 2.55\, M_\odot$ we also analyze a subset of models without neutrino absorption in order to isolate its effect on the ejecta mass and properties, $r$-process yields and kilonova emission. This subset is evolved only at LR and includes $M255_{00}^{*}$, $M255_{\downarrow^{0.4}0}^{*}$, $M255_{\uparrow^{0.4} 0}^{*}$, $M255_{\downarrow^{0.4} \downarrow^{0.4}}^{*}$ and $M255_{\uparrow^{0.4} \uparrow^{0.4}}^{*}$. For the $M_{\mathrm{tot}} = 2.55\,M_{\odot}$ models, we perform both low-resolution (LR) and high-resolution (HR) simulations, with grid spacings of approximately $308\,\mathrm{m}$ and $222\,\mathrm{m}$, respectively. The remaining models are evolved at LR. Unless otherwise stated, results are shown for the LR simulations.

\textbf{Naming and spin conventions: } Model names include the total mass, mass ratio (for unequal mass models) and spin configuration following~\cite{effofspin}. The spin orientations are defined relative to the orbital angular momentum and include irrotational $(0 0)$, aligned $(\uparrow \uparrow)$, single-spin aligned $(\uparrow 0)$, anti-aligned $(\downarrow \downarrow)$, single-spin anti-aligned $(\downarrow 0)$ and mixed $(\downarrow \uparrow)$ configurations. For example, $M305q205_{\uparrow^{0.6}0}$ denotes a binary with a total mass of $3.05\,M_\odot$ and a mass ratio of $q=2.05$, in which the primary NS has a dimensionless spin of $\chi_1=0.6$ aligned with the orbital angular momentum, while the secondary is irrotational.

\textbf{Ejecta:} Ejecta properties are extracted from a surface located at ${\sim}295\, \mathrm{km}$ from the center of mass. The dynamical ejecta are identified using the geodesic criterion $u_t < -1$, where $u_t$ is the time component of the four-velocity~\citep{kastaun_geodesic_2015, sekiguchi_geodesic_2015, dietrich_geodesic_2015, radice_massejection_2016, radice_massejection_2018, Foucart_ejecta21}.
The dynamical ejecta are further subdivided following~\cite{radice_massejection_2016, radice_massejection_2018} into a tidal component, characterised by low entropy $s < 10\, \mathrm{k_{B}\, baryon^{-1}}$ and predominantly confined to the orbital plane, and a shock-heated component with entropy $s \gtrsim 10\, \mathrm{k_{B}\, baryon^{-1}}$ generated by shocks at and after the merger. 

We also analyze the high-velocity component of the dynamical ejecta, defined by $v > 0.3\, c$ and motivated by the inferred velocity of the blue component of \texttt{AT2017gfo}~\citep{at2017gfo_optik_dlt, at2017gfo_nir}. We define the fast-ejecta component by $v > 0.6\, c$, following~\citet{radice_massejection_2018}. For both components, we examine the ejecta properties, their contribution to the nucleosynthesis yields, and their impact on the resulting kilonova.

We define the secular component using the Bernoulli criterion $h u_t < -1$, where $h$ is the specific enthalpy, and measure it after the dynamical ejecta mass has saturated. Models producing a long-lived NS remnant are evolved to ${\sim}20$ ms after the merger, whereas models that collapse to a BH are evolved for ${\sim}10$ ms after BH formation. At these times, the cumulative secular ejecta mass is still increasing in some models, making the reported secular masses and corresponding nucleosynthesis yields lower limits. Long term simulations show that the secular mass ejection can persist beyond the timescales considered here before saturation is reached~\citep{Radice_secular_24}. These outflows include contributions from neutrino-driven ejecta, which are concentrated at high-latitudes ($\theta < 45^\circ$, where $\theta$ is measured from the polar axis, with $\theta = 0^\circ$ at the pole and $\theta = 90^\circ$ at the orbital plane) with electron fraction $Y_e > 0.25$, and spiral-wind driven ejecta, typically associated with remnant-disc interactions and disc oscillations~\citep{radice_massejection_2016, radice_massejection_2018, spiralwind_nedora, Matur_signatures, matur_bhns_spin_2025}.

\textbf{Nucleosynthesis:} The ejecta properties are used as input to the \texttt{SkyNet}~\citep{skynet, skynet_paper, skynet_code} nuclear reaction network to compute the $r$-process nucleosynthesis yields. For a mass number interval $[A_1,A_2)$, we define the yield as 
\begin{equation}
     M_{[A_1, A_2)} = \sum_i m_i \sum_{A_1 \leq A < A_2} A\, Y_i(A)
\end{equation}
where $m_i$ is the mass associated with ejecta bin $i$ and $Y_i(A)$ is its final number abundance. Normalized abundance patterns are used only to compare the relative distribution in $A$. In the following, the mass fraction at $A\geq140$, $X_{A\geq140} = M_{A\geq140}/M_\mathrm{ej}$, is used to quantify the relative production of heavy $r$-process nuclei, while the mass in this range, $M_{A\geq140}$, gives the absolute yield. Here, $M_\mathrm{ej}$ denotes the mass of the ejecta component under consideration. Elemental yields are obtained by grouping the same mass-weighted isotopic contributions by atomic number $Z$. We define the lanthanide mass fraction over $57 \leq Z \leq71$ and the light $r$-process mass fraction from Sr, Y, and Zr and use their ratio to characterize the relative production of lanthanides and light $r$-process elements.

Each ejecta bin is assigned a parametrized nucleosynthesis trajectory specified by the initial electron fraction $Y_e$, specific entropy $s$, and expansion timescale $\tau$~\citep{skynet}. We investigate the contributions of the dynamical ejecta as a whole, its tidal, shock-heated, high-velocity and fast subsets, and the secular ejecta to the $r$-process yields. The electron fraction, specific entropy, velocity and rest-mass density are extracted at the outflow surface. Following~\citet{radice_massejection_2016, radice_massejection_2018}, homologous expansion is assumed beyond this radius and the velocity and rest-mass density are used to estimate $\tau$. 
The network calculation is initialized at a temperature of ${\sim}6\, \mathrm{GK}$. Full details of the procedure are given in~\cite{radice_massejection_2016, radice_massejection_2018}. The validity of this approach was assessed by~\citet{radice_massejection_2018} by comparison with nucleosynthesis calculations based on tracer particles. The resulting abundance distributions were found to agree within a factor of two at all mass numbers. The solar $r$-process abundance pattern used for comparison is taken from~\cite{Arlandini_1999}.

\textbf{Kilonova:} We compute kilonova light curves with the semi-analytic model \texttt{xkn}~\citep{xkn_2024}. For each simulation-based ejecta component and angular bin, we use the ejecta mass, mass-weighted velocity, electron fraction and entropy extracted from the simulation profiles. We extend the public implementation of \texttt{xkn} to include the simulation-derived angular entropy, $s(\theta)$, and expansion timescale, $\tau(\theta)$, in the radioactive heating calculation for each angular bin. To isolate the effect of this extension, we also compute baseline models with identical ejecta mass, velocity and electron fraction profiles. For each ejecta component, the baseline adopts its mass-weighted mean entropy and sets $\tau=1/v$, while the extended models use the angular profiles $s(\theta)$ and $\tau(\theta)$.
 The composition-dependent effective grey opacity is computed in each angular bin from the simulation derived electron fraction profile, $Y_e(\theta)$, using the $\kappa(Y_e)$ prescription implemented in \texttt{xkn}. We therefore retain the continuous angular composition structure and do not impose discrete red, purple and blue opacity classes.

To assess the sensitivity of the light curves to the treatment of long-term disc outflows, we also compute a separate hybrid model. In this model, the simulation based dynamical ejecta are combined with a parametric disc outflow prescription instead of the simulation based secular profile. For the hybrid calculations, we use the disc masses reported in our previous study~\citep{effofspin}. The disc mass is computed by integrating the mass in regions with $\rho < 10^{13}\,\mathrm{g\,cm^{-3}}$ within a radius of ${\sim}295$ km. For models forming a BH, regions with lapse $\alpha < 0.3$ are excluded. Motivated by the magnetohydrodynamic results of~\citet{Siegel_disc_2018}, we adopt a parametric disc outflow prescription corresponding to $40$ per cent of the disc mass, with $Y_e = 0.20$ and $v=0.10\,c$. The simulation based secular ejecta and the parametric disc outflow of~\citet{Siegel_disc_2018} are not added simultaneously.

For the three fixed $\chi_\mathrm{eff}=0$ models with $M_\mathrm{tot} = 2.55\,M_\odot$, we test whether the spin imprint persists when the simulation derived secular ejecta are replaced by the same parametric disc outflow prescription. The prescription is applied to disc masses of $0.19$, $0.16$ and $0.23\,M_\odot$ for models $M255_{00}$, $M255_{\downarrow^{0.4} \uparrow^{0.4}}$ and $M255_{\downarrow^{0.65} \uparrow^{0.65}}$, respectively.

We evaluate the model emission at representative wavelengths of $482.6$, $754.5$ and $2148.6\, \mathrm{nm}$, chosen to represent the $g$, $i$ and $K_s$ bands, respectively. For each viewing angle, we determine the peak absolute magnitude and corresponding peak time from the minimum of each light curve, using local quadratic refinement in $\log_{10}t$ around the minimum. We define the colours as $g-i$ and $i-K_s$ from magnitudes evaluated at the same epoch. For each model pair, the peak associated colour separation is defined as the maximum absolute colour difference over the six epochs corresponding to the $g$, $i$ and $K_s$ peaks of the two models. The peak times are determined from light curves of each model.

For the neutrino absorption comparison, we compute the light curves separately from the corresponding ejecta profiles obtained with and without neutrino absorption, while keeping the kilonova treatment otherwise unchanged. We define $\Delta M_X^\nu = M_{X, \mathrm{pk}}^\mathrm{no\,abs} - M_{X, \mathrm{pk}}^\mathrm{abs}$, such that $\Delta M_X^\nu>0$ corresponds to a brighter peak when neutrino absorption is included. Similarly, we define $\Delta C^\nu = C^\mathrm{no\,abs}- C^\mathrm{abs}$, so that a positive value corresponds to a blueward colour shift when neutrino absorption is included.

We consider viewing angles from $0^\circ$ to $90^\circ$ in steps of $10^\circ$, with $\theta_\mathrm{obs}=0^\circ$ and $\theta_\mathrm{obs}=90^\circ$ corresponding to the polar and equatorial viewing angles, respectively. For observational reference, we convert the model absolute magnitudes to apparent magnitudes at $40\, \mathrm{Mpc}$ and compare them with representative $5\sigma$ point-source limiting magnitudes of $g_\mathrm{AB} = 24.5$ and $i_\mathrm{AB} = 23.4$ for a single $30\,\mathrm{s}$ Rubin Observatory exposure~\citep{Rubin}, and $K_{s, \mathrm{AB}} = 24.2$ for a $1\, \mathrm{h}$ VLT/HAWK-I integration~\citep{Hawk}. For the observability check, we define an epoch as jointly observable in a given band when both models are brighter than the adopted $5\sigma$ limiting magnitude at the same epoch. For a colour, both constituent bands are required to satisfy this condition for both models. The viewing angle variation $\Delta_\theta M_\mathrm{pk}$ is the difference between the maximum and minimum peak absolute magnitudes over this range. Unless stated otherwise, peak magnitudes are reported as absolute AB magnitudes, while apparent magnitudes at $40\,\mathrm{Mpc}$ are used only for the observational comparison.

\section{Ejecta}\label{sec:ejecta}

\begin{figure*}
    \centering
    \includegraphics[width=\textwidth]{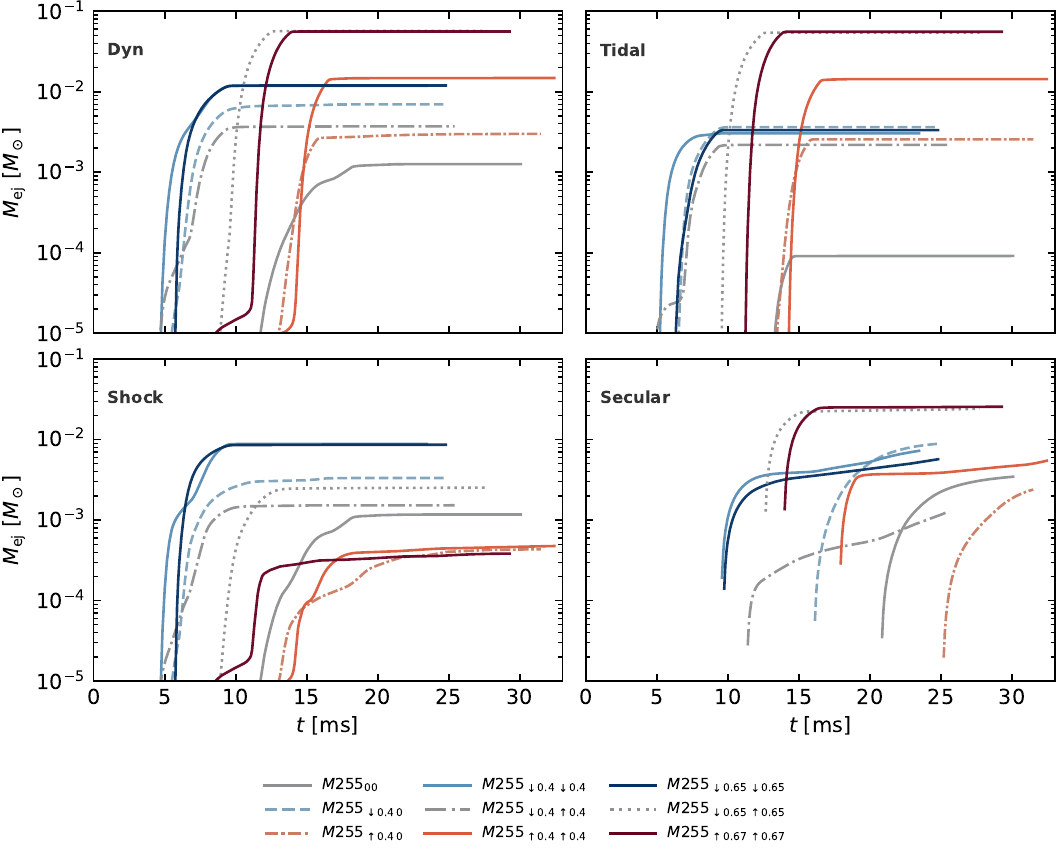}
    \caption{Cumulative ejecta masses for the $M_\mathrm{tot} = 2.55\,M_\odot$ models, shown for the dynamical, tidal, shock-heated and secular ejecta components.}
    \label{fig:ejecta_mass}
\end{figure*}

We first examine how spin changes the dynamical ejecta mass and the relative contributions of tidal and shock-heated material, and the high-velocity and fast ejecta, and then assess how these differences are reflected in the composition, thermodynamic, angular and velocity properties of the ejecta. We subsequently consider the secular ejecta and the effects of neutrino absorption. The ejecta masses are summarized in Table~\ref{tab:ejecta_masses} and Fig.~\ref{fig:ejecta_mass}, while the ejecta properties and the effects of neutrino absorption are presented in Fig.~\ref{fig:ejecta_properties} and Fig.~\ref{fig:ejecta_neutrino}.

\subsection{Dynamical ejecta}\label{sec:dynamical_ejecta}

\textbf{Mass and ejection channels:} Spin changes both the dynamical ejecta mass and the relative contributions of the tidal and shock-heated components.

For $M_\mathrm{tot} = 2.55 M_\odot$, the spinning models eject $(3.0 -57)\times10^{-3}\,M_\odot$ of dynamical material. The irrotational model is shock dominated, with ${\sim}93$ per cent of the dynamical ejecta in the shock-heated component. The single-spin anti-aligned model $M255_{\downarrow^{0.4}\, 0}$, in which the spinning NS is tidally disrupted, shows comparable tidal and shock-heated contributions. Anti-aligned models remain shock-dominated, but with a substantial tidal contribution of ${\sim}25$-$30$ per cent. Aligned, single-spin aligned and mixed configurations are tidally dominated, with the tidal fraction reaching ${\sim}99$ per cent in $M255_{\uparrow^{0.67} \uparrow^{0.67}}$. 

At fixed total mass, mass ratio, EOS and $\chi_\mathrm{eff} = 0$, the irrotational and mixed-spin models span a factor of ${\sim}45$ in dynamical ejecta mass, while their tidal fractions range from ${\sim}7$ to ${\sim}96$ per cent. These differences demonstrate that $\chi_\mathrm{eff}$ alone does not uniquely determine either the dynamical ejecta mass or its decomposition into tidal and shock-heated components. The relative contributions of tidal and shock-heated ejecta reflect the spin dependent merger dynamics identified in our previous analysis~\citep{effofspin}. Relative to the irrotational model, anti-aligned models have lower total angular momentum and undergo stronger shock heating, whereas aligned and mixed models favour tidal mass ejection.

The $M_\mathrm{tot} = 3.05 M_\odot$ models provide controlled comparisons of the effect of mass ratio at fixed $\chi_\mathrm{eff}$. At $\chi_\mathrm{eff}=0$, increasing the mass ratio from $q=1$ to $q=2.05$ increases the dynamical ejecta mass by a factor of ${\sim}30$ and changes the ejecta from entirely shock-heated to tidal dominated, although both models promptly collapse to BHs. At fixed $\chi_\mathrm{eff} = 0.4$, the equal mass and unequal mass models, $M305_{\uparrow^{0.4}\uparrow^{0.4}}$ and $M305q205_{\uparrow^{0.6} {0}}$, are strongly tidal dominated, while the unequal mass model ejects $\sim2$ times as much dynamical material. Within the equal mass models, the anti-aligned model $M305_{\downarrow^{0.4}\downarrow^{0.4}}$ preserves shock dominance, whereas aligned spin reverses the channel balance and increases the dynamical ejecta mass by factors of ${\sim}20$ and ${\sim}150$ for $M305_{\uparrow^{0.4}\uparrow^{0.4}}$ and $M305_{\uparrow^{0.67}\uparrow^{0.67}}$, respectively, relative to $M305_{00}$. The fixed $\chi_\mathrm{eff}$ comparisons show that the mass ratio can change both the ejecta mass and the dominant ejection channel, whereas at $\chi_\mathrm{eff}=0.4$, where both models are tidal dominated, its main effect is on the ejecta mass.

Although all $M_\mathrm{tot} = 4.10 M_\odot$ models undergo prompt collapse, their dynamical ejecta remain strongly dependent on the spin configuration. The irrotational model produces only ${\sim}2\times10^{-5}\,M_\odot$ of dynamical ejecta, which are almost entirely shock-heated, while the anti-aligned model produces neither ejecta nor a disc. At $\chi_\mathrm{eff}=0$, the mixed model produces nearly three orders of magnitude more dynamical ejecta mass than the irrotational model and is tidal dominated. The aligned model likewise produces substantial tidal-dominated ejecta despite undergoing prompt collapse.

Previous irrotational studies found that increasing mass asymmetry generally enhances the tidal contribution to the dynamical ejecta, while the total dynamical ejecta mass remains sensitive to the total mass and EOS~\citep{sekiguchi_massratio:2016,dietrich_massratio2017, radice_massejection_2018}. Predominantly tidal ejecta were also reported for an irrotational $q=2.06$ model evolved with a different EOS treatment~\citep{dietrich_q206_2015}, while previous spinning BNS studies found that the dynamical ejecta depend on binary and spin configuration~\citep{dietrich_spin_2017, East_spin_2019}. These trends are consistent with the mass ratio and spin dependence seen in our results.

\begin{figure*}
    \centering
    \includegraphics[width=\textwidth]{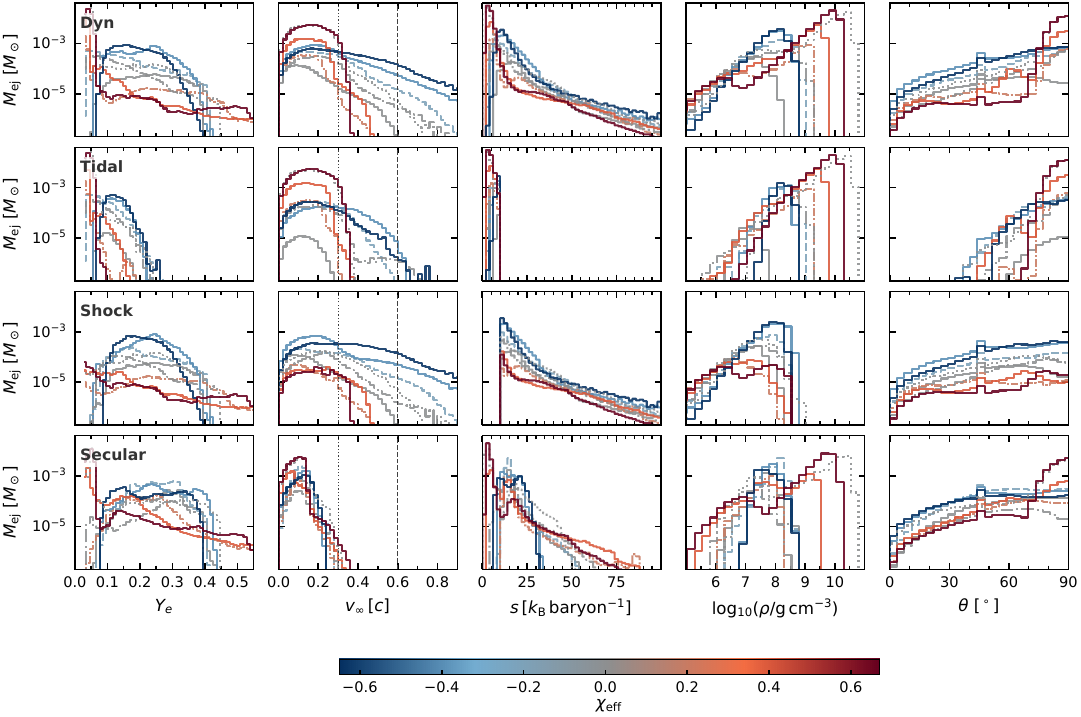}
    \caption{Variation of ejecta properties with $\chi_\mathrm{eff}$ for the $M_\mathrm{tot} = 2.55\,M_\odot$ models. The colours indicate $\chi_\mathrm{eff}$, while models with the same $\chi_\mathrm{eff}$ are distinguished using the same linestyles as in Fig.\ref{fig:ejecta_mass}. Rows show the dynamical, tidal, shock-heated, and secular ejecta, while columns show the electron fraction $Y_e$, asymptotic velocity $v_\infty$, entropy per baryon $s$, rest-mass density $\rho$, and polar angle $\theta$. In the velocity column, the vertical lines at $v_\infty=0.3\,c$ and $0.6\,c$ mark the thresholds used to define the high-velocity and fast ejecta, respectively.}
     \label{fig:ejecta_properties}
\end{figure*}

\textbf{Composition, thermodynamic and angular properties:} The differences in the tidal and shock-heated contributions identified above are reflected in the composition, entropy, density and angular distribution of the dynamical ejecta as shown in~\cref{fig:ejecta_properties}.

For $M_\mathrm{tot} = 2.55\,M_\odot$, tidal-dominated models are concentrated near the orbital plane, whereas shock-dominated models show broader angular distributions, consistent with previous irrotational studies~\citep{Hotokezeka_kilonova_2013, radice_massejection_2016, radice_massejection_2018}. The dynamical ejecta from the irrotational model peak at ${\simeq}20^\circ{-}30^\circ$ from the orbital plane, whereas the aligned, single-spin aligned and mixed-spin models peak within $10^\circ$ of the orbital plane. The stronger equatorial concentration of the ejecta in the aligned, single-spin aligned and mixed-spin models tracks tidal dominance rather than a direct monotonic dependence on spin magnitude. The same channel dependence extends to the thermodynamic properties. The tidal ejecta remain neutron rich and low entropy across all models, with nearly all of their mass at $Y_e < 0.25$. Within this regime, the aligned, single-spin aligned and mixed-spin models have lower $Y_e$ than the anti-aligned models. The shock-heated component is hotter and less neutron rich than the tidal ejecta and produces the high $Y_e$ dynamical ejecta. The higher $Y_e$ is consistent with stronger weak interaction processing in the shock-heated material. The dominant ejection channel is also reflected in the mass-weighted density of the total dynamical ejecta, which is up to three orders of magnitude larger in tidal dominated models than in shock-dominated models.

At fixed $\chi_\mathrm{eff} = 0.4$, the equal mass and unequal mass models, $M305_{\uparrow^{0.4} \uparrow^{0.4}}$ and $M305q205_{\uparrow^{0.6} {0}}$, have similarly neutron-rich, low-entropy and equatorially concentrated dynamical ejecta, despite their different ejecta masses. In the unequal mass models, aligned spin changes the remnant fate from prompt BH formation to a long-lived NS remnant while preserving the neutron-rich, low entropy and equatorially concentrated signature of tidal ejection.

The channel-based interpretation also holds for $M_\mathrm{tot}=4.10 M_\odot$ prompt collapse models that produce ejecta. The tidal components of the mixed and aligned models have nearly identical composition, entropy, density and angular concentration despite their different ejecta masses. The low mass ejecta from the irrotational model instead exhibit the hotter, less neutron-rich and more broadly distributed signature of shock-heated material.

\textbf{Velocity, high-velocity and fast ejecta:} For $M_\mathrm{tot} = 2.55\,M_\odot$, unlike the composition, entropy, density and angular distribution, the ejecta velocity is not determined by the dominant ejection channel alone. The irrotational model is almost entirely shock-heated, but has only $\langle v \rangle \sim0.14\,c$. The highest velocities occur in the anti-aligned models, with $M255_{\downarrow^{0.65} \downarrow^{0.65}}$ reaching $\langle v \rangle {\sim}0.30\,c$ for the dynamical ejecta and ${\sim}0.35\,c$ for the shock-heated component. By contrast, the aligned and mixed-spin models with the largest ejecta masses retain lower mean velocities, consistent with their predominantly tidal origin. These contrasts show that neither the shock-heated fraction nor the dynamical ejecta mass alone determines the velocity, which remains sensitive to the individual spin configuration and the associated merger dynamics identified in our previous analysis~\citep{effofspin}.

The fraction of the dynamical ejecta with $v>0.3\,c$ increases from ${\sim}7$ per cent in the irrotational model to ${\sim}25$ and ${\sim}40$ per cent in the anti-aligned models with $\chi_\mathrm{eff}=-0.4$ and $-0.65$, respectively. By contrast, the high spin aligned and mixed models produce the largest dynamical ejecta masses but place only a few per cent of this material above $0.3\,c$. The high-velocity fraction is not systematically related to either the total dynamical ejecta mass or the tidal fraction.

At fixed $\chi_\mathrm{eff}=0$, the mass above $0.3\,c$ spans a factor of ${\sim}20$. However, the high-velocity fraction peaks in the lower spin mixed model and decreases in the higher spin mixed model because the total dynamical ejecta mass increases more rapidly than the mass above $0.3\,c$. The resolution dependence of these high-velocity ejecta quantities is discussed in Sec.~\ref{sec:resolution_dependence}. Within the same $\chi_\mathrm{eff}=0$ comparison, the irrotational model produces higher entropy, less neutron rich, lower density and more broadly distributed high-velocity ejecta, whereas the mixed-spin models produce more neutron rich, denser and more equatorially concentrated material. 
The same $\chi_\mathrm{eff}$ can therefore correspond to high-velocity ejecta with very different masses, thermodynamic properties and angular distributions.
The irrotational and aligned spin models show no or only negligible ejecta with $v>0.6\,c$, whereas the anti-aligned models produce the largest fast-ejecta masses. In $M255_{\downarrow^{0.65} \downarrow^{0.65}}$, the mass above $0.6\,c$ reaches ${\sim}10^{-3}\,M_\odot$, with a mean velocity of ${\sim}0.72\,c$. The mixed-spin models with $\chi_\mathrm{eff}=0$ also produce ejecta with $v>0.6\,c$, but in substantially smaller amounts than the high spin anti-aligned model. The $v> 0.3\,c$ ejecta can exhibit either tidal or shock-heated properties, whereas the $v>0.6\,c$ ejecta consistently have high entropy and are less equatorially concentrated, consistent with shock-heated material.

For $M_\mathrm{tot} = 3.05\, M_\odot$, the mean dynamical ejecta velocity distinguishes the shock-dominated and tidal-dominated models. The shock-dominated irrotational equal mass model has the highest dynamical ejecta velocity, $\langle v \rangle {\sim}0.47\,c$, whereas the tidal-dominated models have lower velocities of ${\sim}0.13$-$0.19\,c$. In the equal mass models, the shock-dominated models have the largest high-velocity fractions, whereas aligned spin shifts most of the ejecta into a more massive and slower tidal component.

The high-velocity ejecta in the irrotational and anti-aligned equal mass models have higher entropy and higher $Y_e$ than those in the aligned models, while the high spin aligned model is much more strongly concentrated near the orbital plane. Despite their similar high-velocity fractions, $M305_{\uparrow^{0.67} \uparrow^{0.67}}$ and $M305q205_{00}$ produce physically distinct high-velocity ejecta. The ejecta from $M305_{\uparrow^{0.67} \uparrow^{0.67}}$ are neutron rich, low entropy, dense and concentrated near the orbital plane, whereas those from $M305q205_{00}$ are less neutron rich, high entropy, lower density and more broadly distributed. Their fast ejecta masses also differ strongly, with $M305q205_{00}$ producing ${\sim}3\times10^{-4}\,M_\odot$, the largest fast ejecta mass among the $M_\mathrm{tot} = 3.05\,M_\odot$ models, while $M305_{\uparrow^{0.67} \uparrow^{0.67}}$ produces no fast ejecta.

At fixed $\chi_\mathrm{eff}=0.4$, the equal and unequal mass models are tidal dominated, with the unequal mass model ejecting more than twice as much tidal ejecta, but their high-velocity fractions differ by a factor of $7$. Their high-velocity ejecta also have different thermodynamic and angular properties, and only the unequal mass model $M305q205_{\uparrow^{0.6} {0}}$ produces fast ejecta. Effective spin, dominant ejection channel and high-velocity fraction are individually insufficient to determine the thermodynamic and angular properties of the high-velocity ejecta or the fast ejecta mass.

For $M_\mathrm{tot} = 4.10\, M_\odot$, the three models that produce ejecta have comparable dynamical ejecta velocities of ${\sim}0.20\,c$ despite their different dominant ejection channels. Their high-velocity ejecta masses span more than three orders of magnitude, while the corresponding fractions remain within ${\sim}20$--$30$ per cent. At fixed $\chi_\mathrm{eff}=0$, the mixed-spin model produces a factor of ${\sim}500$ more high-velocity ejecta than the irrotational model. The high-velocity ejecta from the irrotational model have high entropy, higher $Y_e$, lower density, and a broader angular distribution, whereas the mixed-spin ejecta are neutron rich, low entropy, dense and concentrated near the orbital plane. Despite large differences in their masses and physical properties, the high-velocity ejecta in all three models have nearly identical velocities of ${\sim}0.36\,c$. The aligned model produces about four times more high-velocity ejecta than the mixed-spin model. By contrast, the mixed-spin model produces ${\sim}6\times10^{-5}\,M_\odot$ of fast ejecta, about a factor of $60$ more than the aligned model. The fast ejecta in both spinning models have high entropy, higher $Y_e$ and nearly identical angular distributions, despite their substantially different masses.

\begin{table*}
\centering
\caption{Ejecta properties for all models. The columns give the effective spin, $\chi_\mathrm{eff}$, the masses of the total dynamical ejecta and its tidal, shock-heated, high-velocity ($v>0.3\,c$),  fast ($v>0.6\,c$) and the secular ejecta identified using the Bernoulli criterion ($h u_t < -1$), along with the mass-weighted electron fraction and velocity of the dynamical ejecta. Models marked with $^{*}$ do not include neutrino absorption and are evolved only at LR. For models evolved at both resolutions, values in parentheses correspond to the LR simulations, while the unparenthesized values correspond to the HR simulations.}

\label{tab:ejecta_masses}
\begin{tabular}{lccccccccc}
\hline

Model & $\chi_\mathrm{eff}$ & $M_\mathrm{dyn}$ & $M_\mathrm{tidal}$ & $M_\mathrm{shock}$ & $M_{\mathrm{dyn}}^{v>0.3\,c}$ & $M_{\mathrm{dyn}}^{v>0.6\,c}$ & $M_\mathrm{sec}$ & $\langle Y_e \rangle_\mathrm{dyn}$ & $\langle v \rangle_\mathrm{dyn}$ \\
& & [$10^{-3}\,M_\odot$] & [$10^{-3}\,M_\odot$] & [$10^{-3}\,M_\odot$] & [$10^{-3}\,M_\odot$] & [$10^{-5}\,M_\odot$] & [$10^{-3}\,M_\odot$] & & [$c$] \\
\hline
$M255_{00}$ & 0.00 & 1.86 (1.26) & 0.16 (0.09) & 1.70 (1.17) & 0.12 (0.08) & 0.44 (0.00) & 4.33 (3.46) & 0.24 (0.24) & 0.14 (0.14)
  \\
$M255_{00}^{*}$& 0.00 & 0.54  & 0.23  & 0.31 & 0.06 & 0.00 & 0.81 & 0.14 & 0.17
 \\
$M255_{\downarrow^{0.4}0}$ & -0.20 & 3.78 (6.97) & 1.64 (3.63) & 2.14 (3.34) & 0.69 (1.52) & 2.04 (7.29) & 2.65 (8.86) & 0.16 (0.14) & 0.20 (0.21)   \\
$M255_{\downarrow^{0.4}0}^{*}$ &  -0.20  & 4.65  & 2.94  & 1.71 & 1.42 & 17.21 & 1.15 & 0.12 & 0.25\\
$M255_{\uparrow^{0.4} 0}$ &  0.20  & 2.30 (3.00) & 2.11 (2.57) & 0.19 (0.44) & 0.00 (0.02) & 0.00 (0.00) & 0.63 (2.40) & 0.05 (0.07) & 0.13 (0.13)  \\
$M255_{\uparrow^{0.4} 0}^{*}$ &  0.20  & 2.85  & 2.65  & 0.20 & 0.01 & 0.00 & 0.54 & 0.04 & 0.14\\
$M255_{\downarrow^{0.4} \downarrow^{0.4}}$  &  -0.40 & 6.76 (11.86) & 3.79 (3.05) & 2.96 (8.81) & 2.20 (2.91) & 20.38 (34.35) & 4.27 (7.25) & 0.15 (0.20) & 0.26 (0.24)  \\
$M255_{\downarrow^{0.4} \downarrow^{0.4}}^{*}$  &  -0.40  & 6.73  & 2.87  & 3.86 & 1.61 & 19.04 & 2.21 & 0.15 & 0.22 \\
$M255_{\downarrow^{0.4} \uparrow^{0.4}}$  &  0.00  & 2.38 (3.71) & 1.23 (2.19) & 1.15 (1.53) & 0.23 (0.45) & 0.17 (1.47) & 3.55 (1.26) & 0.14 (0.12) & 0.18 (0.20)  \\
$M255_{\uparrow^{0.4} \uparrow^{0.4}}$  &  0.40  & 11.51 (14.82) & 11.11 (14.34) & 0.40 (0.48) & 0.09 (0.09) & 0.00 (0.00) & 5.68 (5.47) & 0.04 (0.04) & 0.14 (0.14)  \\
$M255_{\uparrow^{0.4} \uparrow^{0.4}}^{*}$ &   0.40  & 13.46  & 13.06  & 0.40 & 0.08 & 0.00 & 5.44 & 0.04 & 0.14  \\
$M255_{\downarrow^{0.65} \downarrow^{0.65}}$  &  -0.65 & 17.33 (11.97) & 10.04 (3.34) & 7.29 (8.63) & 5.48 (5.25) & 59.82 (86.57) & 5.83 (5.70) & 0.14 (0.18) & 0.25 (0.30) \\
$M255_{\downarrow^{0.65} \uparrow^{0.65}}$  &  0.00  & 58.78 (56.83) & 55.50 (54.30) & 3.28 (2.53) & 1.75 (1.50) & 0.24 (1.13) & 2.25 (24.16) & 0.06 (0.05) & 0.17 (0.16)\\
$M255_{\uparrow^{0.67} \uparrow^{0.67}}$&  0.67  & 58.67 (56.08) & 58.38 (55.70) & 0.28 (0.38) & 0.64 (1.30) & 0.00 (0.00) & 28.71 (25.62) & 0.04 (0.04) & 0.16 (0.17) \\
$M305_{00}$ & 0.00  & 0.39  & 0.00  & 0.39 & 0.26 & 11.48 & $5.56\times10^{-3}$ & 0.35 & 0.47  \\
$M305_{\downarrow^{0.4} \downarrow^{0.4}}$ & -0.40  & 0.66  & 0.10  & 0.56 & 0.26 & 4.30 & 0.14 & 0.22 & 0.29  \\
$M305_{\uparrow^{0.4} \uparrow^{0.4}}$ &0.40  & 8.23  & 7.81  & 0.42 & 0.14 & 0.00 & 1.85 & 0.04 & 0.15  \\
$M305_{\uparrow^{0.67} \uparrow^{0.67}}$ &0.67  & 57.98  & 57.41  & 0.56 & 4.43 & 0.00 & 20.66 & 0.04 & 0.19 \\
$M305q205_{00}$ & 0.00  & 10.98  & 9.60  & 1.38 & 0.67 & 29.45 & 0.14 & 0.07 & 0.15 \\
$M305q205_{\uparrow^{0.6} {0}}$ & 0.40  & 17.83  & 16.72  & 1.11 & 0.04 & 0.55 & 0.91 & 0.06 & 0.13 \\
$M410_{00}$ & 0.00  & 0.02  & 0.00  & 0.02 & $6.40\times10^{-3}$ & 0.00 & $2.50\times10^{-3}$ & 0.30 & 0.24 \\
$M410_{\downarrow^{0.65} \uparrow^{0.65}}$ &0.00  & 17.58  & 16.68  & 0.89 & 3.16 & 6.34 & 0.28 & 0.06 & 0.21  \\
$M410_{\downarrow^{0.65} \downarrow^{0.65}}$ &-0.65  & 0.00  & 0.00  & 0.00 & 0.00 & 0.00 & 0.00 & 0.00 & 0.00  \\
$M410_{\uparrow^{0.67} \uparrow^{0.67}}$  & 0.67  & 48.00  & 47.09  & 0.91 & 13.80 & 0.10 & 11.03 & 0.05 & 0.23 \\
\hline
\end{tabular}
\end{table*}

\subsection{Secular ejecta}\label{sec:secular_ejecta}

For $M_\mathrm{tot} = 2.55\,M_\odot$, the high spin mixed and aligned models have the largest cumulative secular ejecta masses, both reaching ${\sim}2.5\times10^{-2}\,M_\odot$ despite their different values of $\chi_\mathrm{eff}$. These models also produce the most neutron-rich, lowest entropy and most equatorially concentrated secular ejecta. The anti-aligned models instead have smaller secular ejecta masses and produce higher $Y_e$, higher entropy and more broadly distributed ejecta. At fixed $\chi_\mathrm{eff} =0$, the secular ejecta mass spans nearly a factor of $20$, accompanied by large differences in the thermodynamic and angular properties. The strong resolution dependence of secular ejecta masses is discussed in Sec.~\ref{sec:resolution_dependence}. The fixed $\chi_\mathrm{eff}$ comparison shows that the individual spin configuration affects both the secular ejecta mass and its thermodynamic and angular properties.

For $M_\mathrm{tot} = 3.05\,M_\odot$, aligned spin increases the secular ejecta mass in both the equal and unequal mass models, but its effect on the ejecta properties depends on the mass ratio. In the equal mass models, the ejecta become more neutron rich, lower entropy and more equatorially concentrated, whereas in the unequal mass models they develop higher $Y_e$, higher entropy and weaker equatorial concentration. 

At fixed $\chi_\mathrm{eff}=0$, $M305_{00}$ and $M305q205_{00}$ differ by a factor of ${\sim}25$ in secular ejecta mass, with the unequal mass model producing lower $Y_e$, lower entropy and more equatorially concentrated ejecta. At fixed $\chi_\mathrm{eff} = 0.4$, $M305_{\uparrow^{0.4} \uparrow^{0.4}}$ and $M305q205_{\uparrow^{0.6} {0}}$ have secular ejecta masses that differ by only a factor of ${\sim}2$, but their composition and angular properties remain distinct. 
These comparisons show that $\chi_\mathrm{eff}$ alone does not uniquely determine the secular ejecta mass and properties across different mass ratios.

For $M_\mathrm{tot} = 4.10\,M_\odot$, the nearly identical thermodynamic and angular properties of the tidal ejecta in the mixed and aligned spin models do not extend to their secular ejecta. The aligned model produces nearly $40$ times more secular ejecta than the mixed-spin model, with lower $Y_e$, lower entropy, higher density and stronger equatorial concentration. The irrotational model produces a much smaller secular component, with the secular ejecta mass spanning nearly four orders of magnitude across the prompt collapse models that produce secular ejecta.

\subsection{Resolution dependence}\label{sec:resolution_dependence}

To assess whether the spin induced differences exceed resolution effects, we compare the $M_\mathrm{tot} = 2.55\,M_\odot$ models evolved at both LR and HR. The dynamical ejecta masses differ by at most a factor of ${\sim}2$ between LR and HR, compared with factors of ${\sim}45$ at LR and ${\sim}30$ at HR among the fixed $\chi_\mathrm{eff}=0$ models. The relative contributions of the tidal and shock-heated ejecta are also resolution dependent, with the anti-aligned models changing from shock-dominated at LR to tidal-dominated at HR. For the high-velocity ejecta, the anti-aligned models have fractions of ${\sim}32$ per cent at HR, while at fixed $\chi_\mathrm{eff}=0$ the mass above $0.3\,c$ spans factors of ${\sim}20$ at LR and ${\sim}15$ at HR. The corresponding heavy $r$-process yield differs by at most a factor of ${\sim}2$ between resolutions, while the fixed $\chi_\mathrm{eff}=0$ models span more than two orders of magnitude at both resolutions. The secular ejecta show a stronger resolution dependence, with mean differences of factors of ${\sim}3$ in mass and ${\sim}4$ in the heavy $r$-process yield, reaching maximum factors of ${\sim}10$ and ${\sim}20$, respectively. These values are lower limits for models in which the cumulative secular ejecta mass has not saturated.

The individual spin imprint in the kilonova persists at both resolutions. The peak magnitudes change by at most ${\sim}0.5$ mag between LR and HR. The $M255_{\downarrow^{0.65} \uparrow^{0.65}}$ model remains the brightest among the fixed $\chi_\mathrm{eff}=0$ models in all bands and at every viewing angle, although the peak magnitude separations change with resolution. The peak associated colour differences also persist at both resolutions, but their values change with resolution.

\subsection{Effect of neutrino absorption}\label{sec:neutrino_absorption}

Neutrino absorption increases the dynamical ejecta mass in all five models, by more than a factor of two in the irrotational model but by at most ${\sim}10$ per cent in the aligned models. The larger increases in the irrotational and anti-aligned models arise mainly from the shock-heated component, while the dynamical ejecta in the aligned models remain tidal dominated. The secular ejecta mass also increases in all models.

\begin{figure*}
    \centering
    \includegraphics[width=\textwidth]{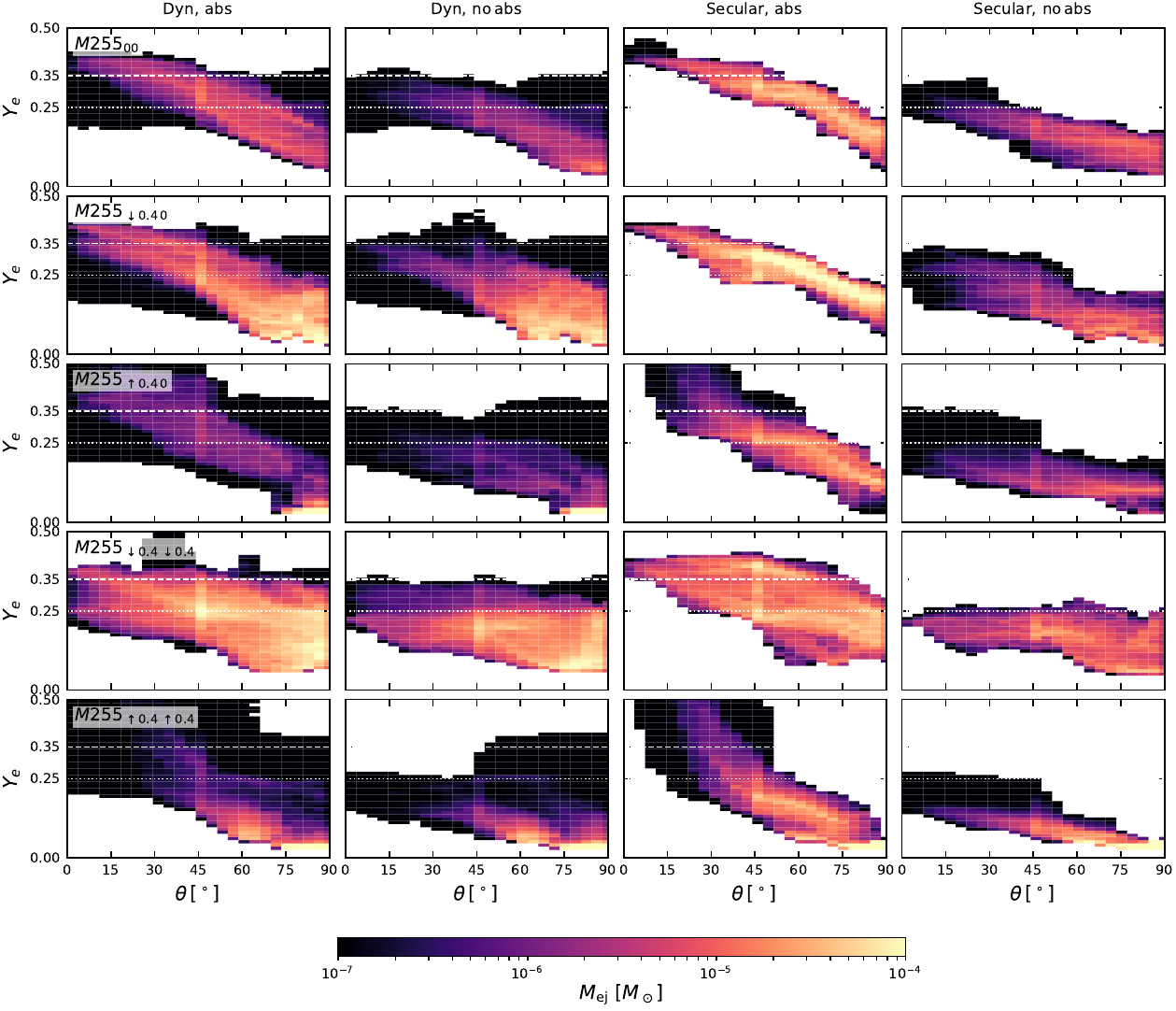}
    \caption{Distribution of the electron fraction of the dynamical and secular ejecta for models with (abs) and without (no abs) neutrino absorption for different spin configurations with $M_\mathrm{tot} = 2.55\,M_\odot$.}
    \label{fig:ejecta_neutrino}
\end{figure*}

Neutrino absorption increases $Y_e$ and the amount of shock-heated ejecta at more than $45^\circ$ above the orbital plane in all models, while the tidal ejecta remain strongly neutron rich and change little. These results extend the behaviour previously identified in irrotational models~\citep{sekiguchi_geodesic_2015, sekiguchi_massratio:2016, radice_massejection_2016, radice_massejection_2018} to spinning configurations. In $M255_{\downarrow^{0.4} \downarrow^{0.4}}$, the fraction of dynamical ejecta with $Y_e < 0.25$ decreases from ${\sim}93$ to ${\sim}60$ per cent, while the fraction with $0.25 \leq Y_e < 0.35$ increases from ${\sim}7$ to ${\sim}40$ per cent. By contrast, nearly all dynamical ejecta in $M255_{\uparrow^{0.4} \uparrow^{0.4}}$ have $Y_e <0.25$ with and without neutrino absorption, as shown in Fig.\ref{fig:ejecta_neutrino}. The high-velocity ejecta mass and $Y_e$ increase in every model when neutrino absorption is included. The largest relative mass change occurs in $M255_{\uparrow^{0.4}0}$, where the high-velocity mass approximately doubles despite only a ${\sim}5$ per cent increase in the total dynamical ejecta mass. The fast ejecta mass does not follow this trend, decreasing by ${\sim}60$ per cent in $M255_{\downarrow^{0.4}0}$ but increasing by ${\sim}80$ per cent in $M255_{\downarrow^{0.4} \downarrow^{0.4}}$, while $Y_e$ increases in both models.

\section{Nucleosynthesis}\label{sec:nucleosynthesis}

\begin{figure*}
    \centering
    \includegraphics[width=\textwidth]{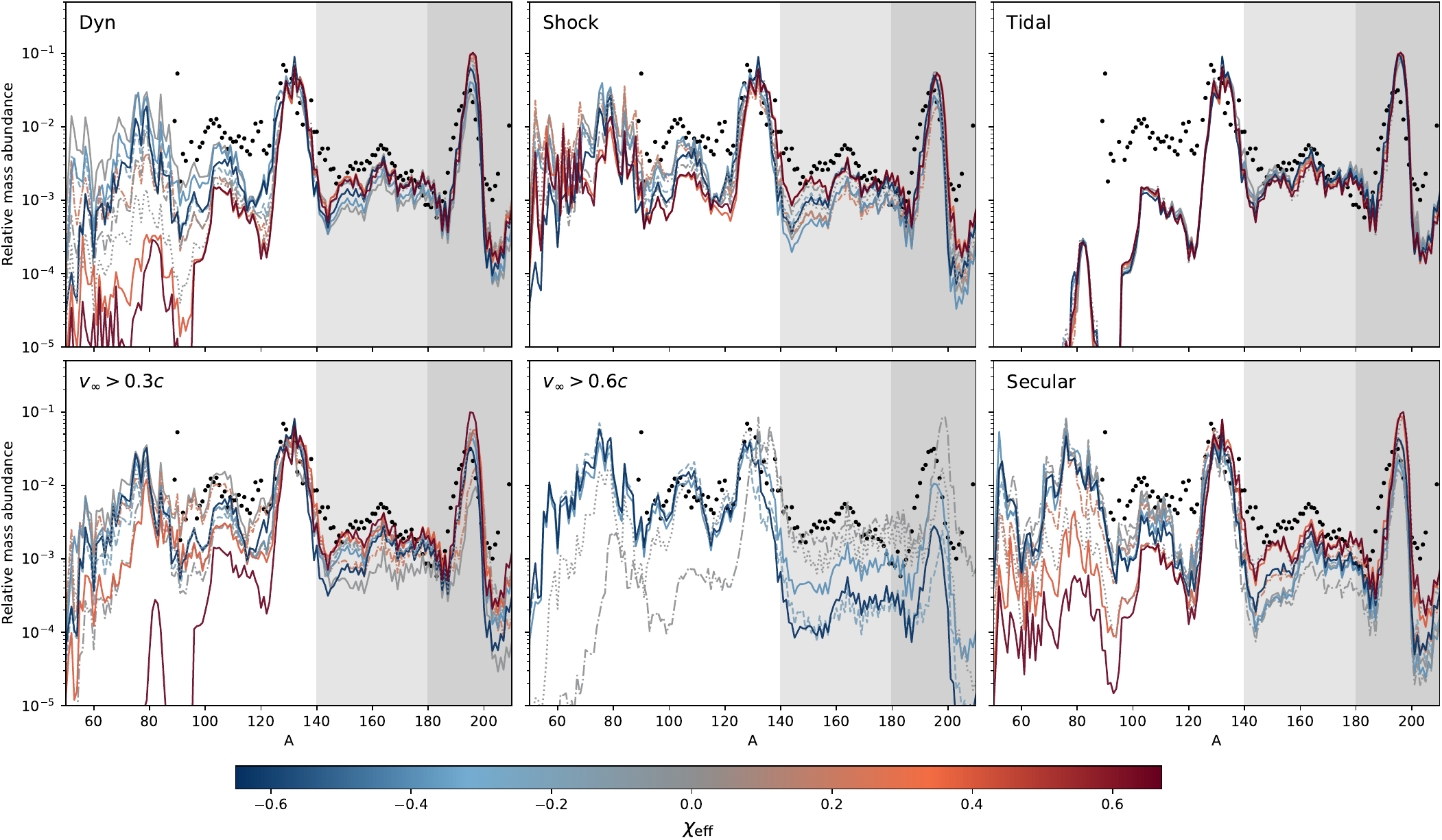}
    \caption{
    Impact of effective spin on the $r$-process nucleosynthesis for $M_\mathrm{tot}=2.55\,M_\odot$ models. Panels show the dynamical, shock-heated, tidal, high-velocity ($v_\infty>0.3\,c$), fast ($v_\infty > 0.6\,c$) and secular ejecta. Black dots show the solar $r$-process abundance pattern, while the shaded regions mark the mass number ranges beyond the second and around the third $r$-process peaks.}
    \label{fig:m255_nucleosynthesis}
\end{figure*}

\begin{figure*}
    \centering
    \includegraphics[width=\textwidth]{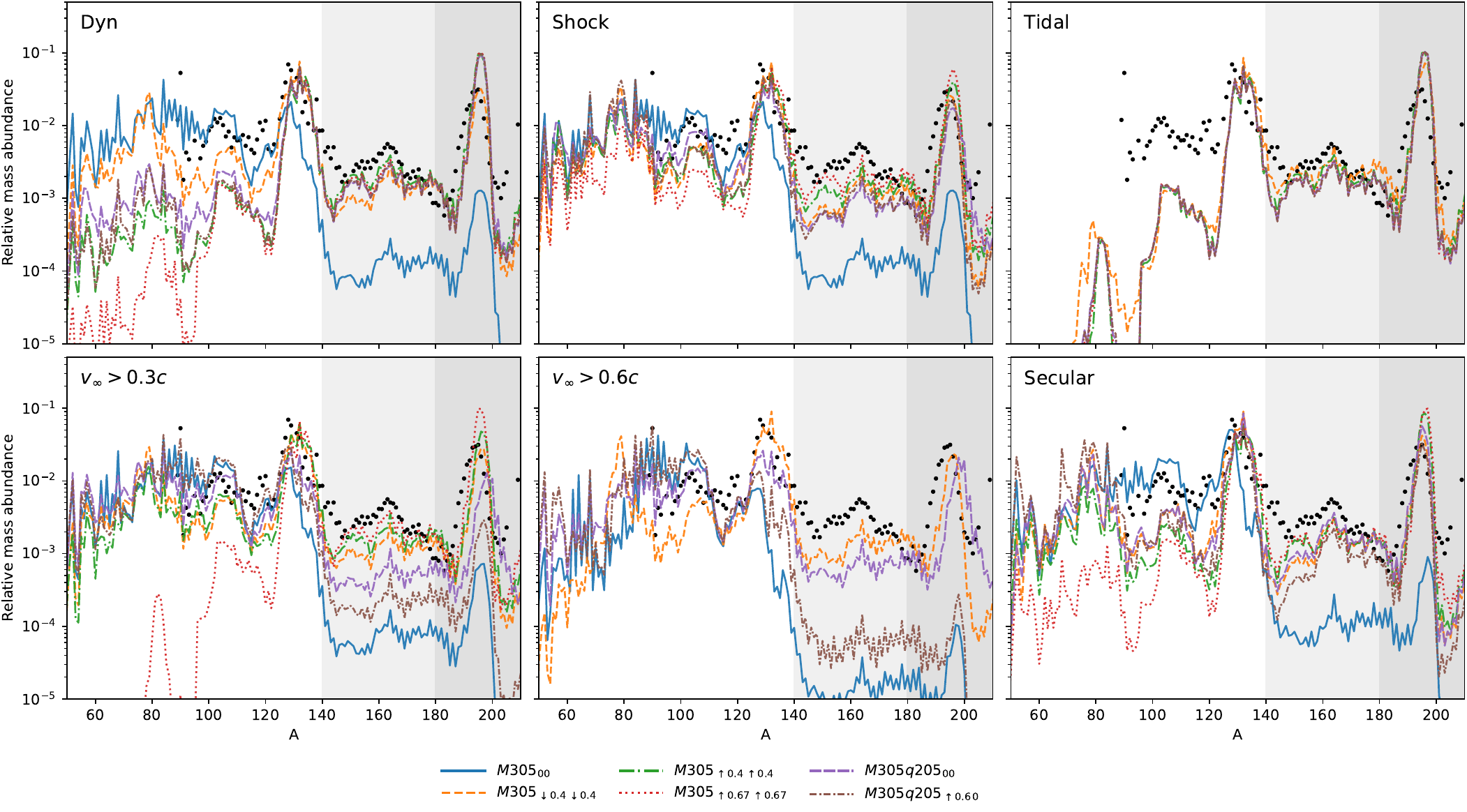}
    \caption{
    Impact of spin and mass ratio on the $r$-process nucleosynthesis for the $M_\mathrm{tot} = 3.05\,M_\odot$ models. The panels, shaded regions and solar abundance pattern are the same as in Fig.~\ref{fig:m255_nucleosynthesis}.
    }
    \label{fig:m305_nucleosynthesis}
\end{figure*}

We investigate how the individual spin configuration, total mass, and mass ratio affect $r$-process nucleosynthesis through changes in the relative contributions and abundance patterns of the ejecta channels. The nucleosynthesis results for $M_\mathrm{tot}= 2.55\,M_\odot$ and $3.05\,M_\odot$ are presented in Fig.\ref{fig:m255_nucleosynthesis} and Fig.\ref{fig:m305_nucleosynthesis}, respectively.

For $M_\mathrm{tot} = 2.55\,M_\odot$, the $r$-process output does not vary monotonically with $\chi_\mathrm{eff}$. The tidal component shows little variation in its heavy $r$-process abundance distribution across the spin configurations, with about $60$ per cent of its mass at $A\geq140$, whereas the shock-heated component varies substantially with spin. The dynamical abundance pattern therefore reflects both the relative contributions of the ejecta channels and spin dependent changes in the abundance distribution of the shock-heated component.

\begin{table}
\centering
\caption{Heavy $r$-process mass fractions, $X_{A\geq140}$, and absolute yields, $M_{A\geq140}$, for the dynamical and secular ejecta of all models. Models marked with $^{*}$ do not include neutrino absorption and are evolved only at LR. For models evolved at both resolutions, values in parentheses correspond to the LR simulations, while the unparenthesized values correspond to the HR simulations.}

\label{tab:r_process}
\footnotesize
\setlength{\tabcolsep}{3pt}
\begin{tabular}{lcccc}
\hline
 &
\multicolumn{2}{c}{Dynamical}& \multicolumn{2}{c}{Secular}\\
Model  & $X_{A\geq140}$ & $M_{A\geq140}$ & $X_{A\geq140}$ & $M_{A\geq140}$ \\

& & [$10^{-3}\,M_\odot$] & & [$10^{-3}\,M_\odot$] \\
\hline

$M255_{00}$  & 0.16 (0.20) & 0.30 (0.26) & 0.12 (0.16) & 0.51 (0.55) \\
$M255_{00}^{*}$& 0.41 & 0.22 & 0.53 & 0.43 \\
$M255_{\downarrow^{0.4}0}$ &  0.42 (0.47) & 1.57 (3.28) & 0.10 (0.14) & 0.27 (1.24)   \\
$M255_{\downarrow^{0.4}0}^{*}$  & 0.53  & 2.45  & 0.57  & 0.65 \\
$M255_{\uparrow^{0.4} 0}$   &  0.59 (0.56) & 1.36 (1.68) & 0.20 (0.25) & 0.12 (0.61)   \\
$M255_{\uparrow^{0.4} 0}^{*}$   & 0.62  & 1.76  & 0.57  & 0.31\\
$M255_{\downarrow^{0.4} \downarrow^{0.4}}$   &  0.46 (0.27) & 3.08 (3.25) & 0.12 (0.13) & 0.52 (0.94)   \\
$M255_{\downarrow^{0.4} \downarrow^{0.4}}^{*}$   & 0.46  & 3.07  & 0.50  & 1.11 \\
$M255_{\downarrow^{0.4} \uparrow^{0.4}}$   &  0.44 (0.53) & 1.05 (1.97) & 0.15 (0.10) & 0.54 (0.13)   \\
$M255_{\uparrow^{0.4} \uparrow^{0.4}}$  &  0.62 (0.63) & 7.18 (9.27) & 0.52 (0.55) & 2.98 (2.99)  \\
$M255_{\uparrow^{0.4} \uparrow^{0.4}}^{*}$ & 0.63  & 8.47  & 0.61  & 3.31 \\
$M255_{\downarrow^{0.65} \downarrow^{0.65}}$  &  0.49 (0.40) & 8.42 (4.76) & 0.40 (0.27) & 2.35 (1.53)  \\
$M255_{\downarrow^{0.65} \uparrow^{0.65}}$    &  0.56 (0.59) & 33.20 (33.30) & 0.28 (0.50) & 0.62 (12.00)  \\
$M255_{\uparrow^{0.67} \uparrow^{0.67}}$   &  0.61 (0.63) & 35.90 (35.10) & 0.56 (0.58) & 16.10 (14.90)  \\
$M305_{00}$ & 0.01  & $5.66\times10^{-3}$  & 0.01  & $0.06\times10^{-3}$  \\
$M305_{\downarrow^{0.4} \downarrow^{0.4}}$   & 0.25  & 0.16  & 0.27  & 0.04  \\
$M305_{\uparrow^{0.4} \uparrow^{0.4}}$& 0.62  & 5.10  & 0.52  & 0.96  \\
$M305_{\uparrow^{0.67} \uparrow^{0.67}}$   & 0.63  & 36.50  & 0.59  & 12.10 \\
$M305q205_{00}$  & 0.57  & 6.28  & 0.34  & 0.05\\
$M305q205_{\uparrow^{0.6} {0}}$ & 0.60  & 10.70  & 0.19  & 0.17 \\
$M410_{00}$   & 0.02  & $0.51\times10^{-3}$  & 0.07  & $0.17\times10^{-3}$\\
$M410_{\downarrow^{0.65} \uparrow^{0.65}}$  & 0.61  & 10.80  & 0.04  & 0.01  \\
$M410_{\downarrow^{0.65} \downarrow^{0.65}}$   & --  & --  & --  & --  \\
$M410_{\uparrow^{0.67} \uparrow^{0.67}}$    & 0.64  & 30.70  & 0.62  & 6.83 \\
\hline
\end{tabular}
\end{table}

The $\chi_\mathrm{eff}=0$ models isolate the role of the individual spins. The heavy $r$-process mass fraction of their dynamical ejecta ranges from ${\sim}20$ to ${\sim}60$ per cent, while the corresponding absolute yield ranges from ${\sim}2.6\times10^{-4}\,M_\odot$ to ${\sim}3.3\times10^{-2}\,M_\odot$. This is the first demonstration that BNS models with the same $\chi_\mathrm{eff}$ can produce heavy $r$-process yields differing by more than two orders of magnitude because the individual spin configurations change the ejecta channel balance and the total ejecta mass. The single-spin aligned model has a larger heavy $r$-process mass fraction, reflecting the larger tidal contribution to its dynamical ejecta, but its absolute yield is about half that of the single-spin anti-aligned model because the total dynamical ejecta mass is smaller. The ratio of the lanthanide mass to the light $r$-process mass in the dynamical ejecta increases from ${\sim}2$ in the irrotational model to ${\sim}20$ and ${\sim}70$ in low and high spin mixed models, respectively.

The anti-aligned models show that the abundance pattern can change even when the ejecta channel masses remain nearly unchanged. Their total dynamical, tidal and shock-heated ejecta masses are similar, but the heavy $r$-process mass fraction of the dynamical ejecta rises from ${\sim}30$ to ${\sim}40$ per cent, while the corresponding fraction in the shock-heated component increases from ${\sim}20$ to ${\sim}30$ per cent. In the aligned models, the dynamical ejecta are almost entirely tidal and the heavy $r$-process mass fraction remains near $60$ per cent, while the absolute yield increases by nearly a factor of four as the dynamical ejecta mass increases.

The nucleosynthesis outcome of the high-velocity and fast ejecta depends on the dominant ejection channel. The high-velocity ejecta  have a heavy $r$-process mass fraction of only ${\sim}20$ per cent in the shock-dominated high spin anti-aligned model, compared with ${\sim}60$ per cent in the tidally dominated high spin aligned model. The anti-aligned models contain most of the fast ejecta mass, but these fast tails remain poor in heavy nuclei. By contrast, the fast tail of $M255_{\downarrow^{0.4} \uparrow^{0.4}}$ is rich in heavy nuclei, but its mass of only ${\sim}10^{-5}\,M_\odot$ makes its contribution to the absolute yield negligible. For the secular ejecta, the high spin mixed and aligned models produce the largest heavy $r$-process yields, of order $10^{-2}\,M_\odot$, because both the ejecta masses and heavy $r$-process mass fractions are large.

Neutrino absorption can shift the dynamical ejecta towards lighter nuclei even as the absolute yield increases. Although the heavy $r$-process mass fraction of the dynamical ejecta decreases in all five models, the absolute yield increases in four because the ejecta mass increases enough to compensate for the lower heavy $r$-process mass fraction. In the irrotational model, this mass fraction is reduced by about half, while the absolute yield increases by ${\sim}15$ per cent. Neutrino absorption leaves the tidal abundance pattern nearly unchanged, but produces substantially larger changes in the shock-heated and high-velocity ejecta.

The dynamical ejecta retain the second and third $r$-process peaks in all models with $M_\mathrm{tot}=2.55\,M_\odot$, while the relative production of lighter nuclei is substantially more sensitive to the ejecta properties, consistent with~\citet{radice_massejection_2018}. Our results further identify the individual spin configuration as an additional factor controlling the light to heavy $r$-process balance through its effect on the relative contributions of the tidal and shock-heated ejecta, even at fixed $\chi_\mathrm{eff} =0$.

For $M_\mathrm{tot} = 3.05\,M_\odot$, the effect of spin on the $r$-process depends strongly on the mass ratio. In the equal mass models, aligned spin changes the dynamical abundance pattern from one with almost no heavy $r$-process material to one dominated by heavy nuclei as the tidal contribution increases. The unequal mass irrotational model is already rich in heavy nuclei because of its substantial tidal component, while aligned spin changes the relative abundance pattern only weakly and increases the absolute yield.

At $\chi_\mathrm{eff} = 0$, increasing the mass ratio from $q=1$ to $q=2.05$ raises the heavy $r$-process mass fraction from below $2$ per cent to ${\sim}60$ per cent and the absolute yield by more than three orders of magnitude. At $\chi_\mathrm{eff} = 0.4$, both models have nearly identical heavy $r$-process mass fractions, while the unequal mass model produces about twice the absolute yield. These comparisons show that the mass ratio can strongly affect the abundance pattern when it changes the dominant ejecta channel, whereas at $\chi_\mathrm{eff} = 0.4$, where both models are tidal dominated, its main effect is on the absolute yield. For the high-velocity ejecta, however, the heavy $r$-process mass fraction is ${\sim}35$ per cent for $q=1$ and only ${\sim}3$ per cent for $q=2.05$. For the secular ejecta, aligned spin increases the heavy $r$-process mass fraction in the equal mass model but decreases it for $q=2.05$. The absolute yield increases in both cases because aligned spin also increases the secular ejecta mass. At $\chi_\mathrm{eff} = 0.4$, the heavy $r$-process mass fraction of the secular ejecta is ${\sim}50$ and $20$ per cent for $q=1$ and $q=2.05$, respectively. 

For $M_\mathrm{tot} = 4.10\, M_\odot$, the irrotational model produces very little heavy $r$-process material, consistent with its almost entirely shock-heated dynamical ejecta. In the mixed-spin and aligned spin models, the dynamical ejecta are instead tidally dominated, with a heavy $r$-process mass fraction of about $60$ per cent. At fixed $\chi_\mathrm{eff}=0$, the irrotational and mixed-spin models both undergo prompt collapse, but their absolute heavy $r$-process yields differ by more than four orders of magnitude. The heavy $r$-process mass fraction of the dynamical ejecta is nearly identical in the mixed and aligned models, but the aligned model has an absolute yield about three times larger because its dynamical ejecta mass is larger.

The heavy $r$-process mass fraction of the secular ejecta is only a few per cent in the mixed model but ${\sim}60$ per cent in the aligned model, while the absolute yields differ by nearly three orders of magnitude. 

\section{Kilonovae}\label{sec:kilonovae}

\begin{figure*}
    \centering
    \includegraphics[width=\textwidth]{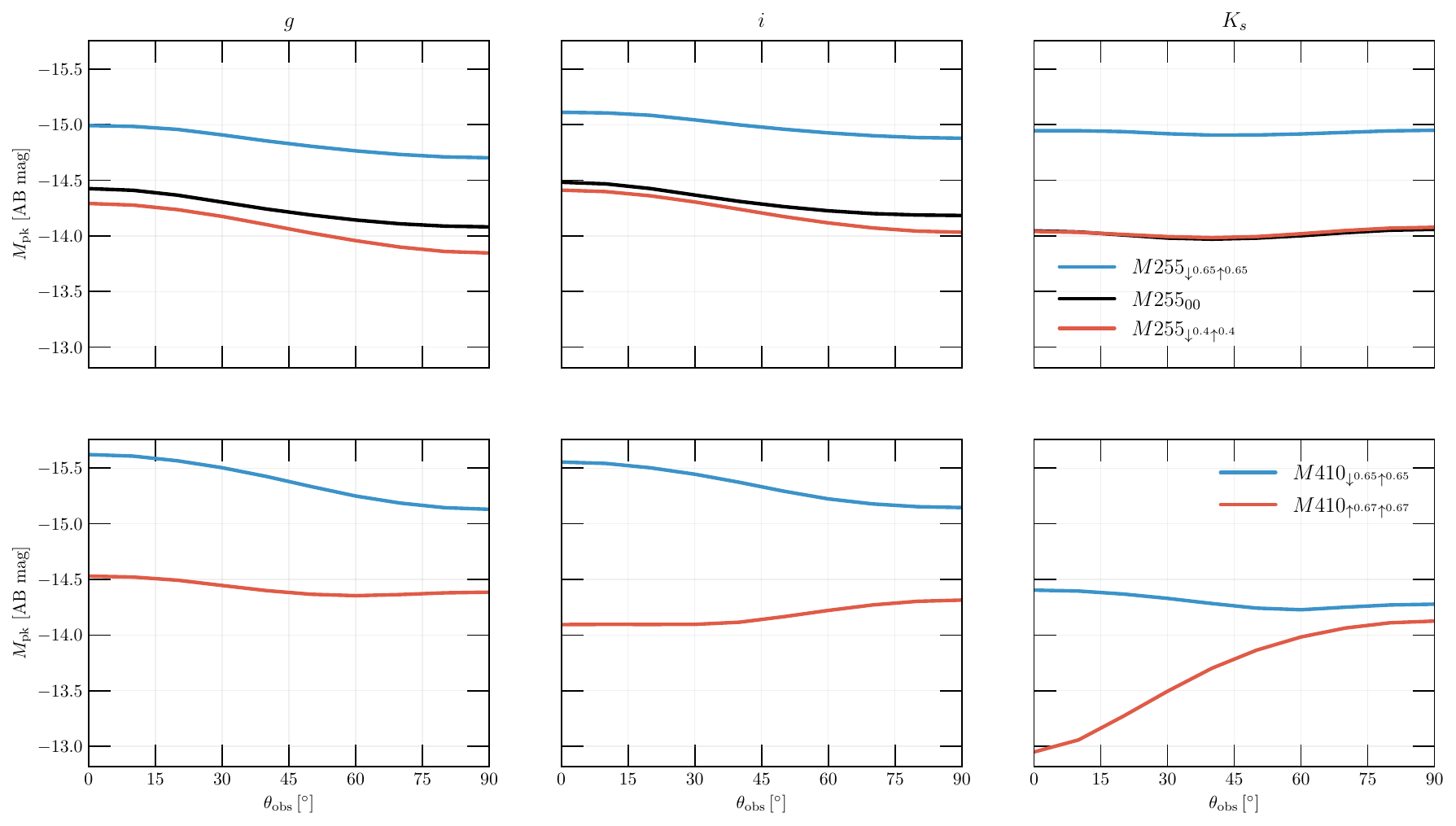}
    \caption{   
    Viewing angle dependence of the kilonova peak absolute magnitudes in the $g$, $i$ and $K_s$ bands. The upper row shows the $M_\mathrm{tot} = 2.55\,M_\odot$ models with $\chi_\mathrm{eff} =0$, while the lower row shows the $M_\mathrm{tot} = 4.10\,M_\odot$ models.}
    \label{fig:kilonova_theta}
\end{figure*}

For $M_\mathrm{tot} = 2.55\,M_\odot$, models with the same effective spin, $\chi_\mathrm{eff}=0$, produce substantially different multi-band kilonova emission. The $M255_{\downarrow^{0.65} \uparrow^{0.65}}$ model is brighter than both the irrotational and $M255_{\downarrow^{0.4} \uparrow^{0.4}}$ models by $\gtrsim0.6$ mag in all three bands for every viewing angle considered, with the largest differences reaching ${\sim}0.9$ mag in $K_s$, as shown in Fig.~\ref{fig:kilonova_theta}. At the epochs corresponding to the band peaks, the fixed $\chi_\mathrm{eff}=0$ models differ by ${\sim}0.2$-$0.7$ mag in $g-i$ and by ${\sim}0.3$-$1.2$ mag in $i-K_s$ across the viewing angle range. Relative to the irrotational model, $M255_{\downarrow^{0.4} \uparrow^{0.4}}$ shows the larger peak associated colour difference, whereas $M255_{\downarrow^{0.65} \uparrow^{0.65}}$ shows the larger peak brightness difference at every viewing angle. The peak brightness ordering therefore does not uniquely determine the colour ordering, showing that the multi band evolution retains information about the individual spin configuration that is not captured by $\chi_\mathrm{eff}$ alone. At $40\, \mathrm{Mpc}$, fixed $\chi_\mathrm{eff}=0$ models retain jointly observable epochs in $g$, $i$, $K_s$, $g-i$, and $i-K_s$ at every viewing angle under the adopted optical and near-infrared depths, with same epoch colour differences reaching ${\sim}1.5$ mag in the simulation based calculation.

\begin{figure*}
    \centering
    \includegraphics[width=\textwidth]{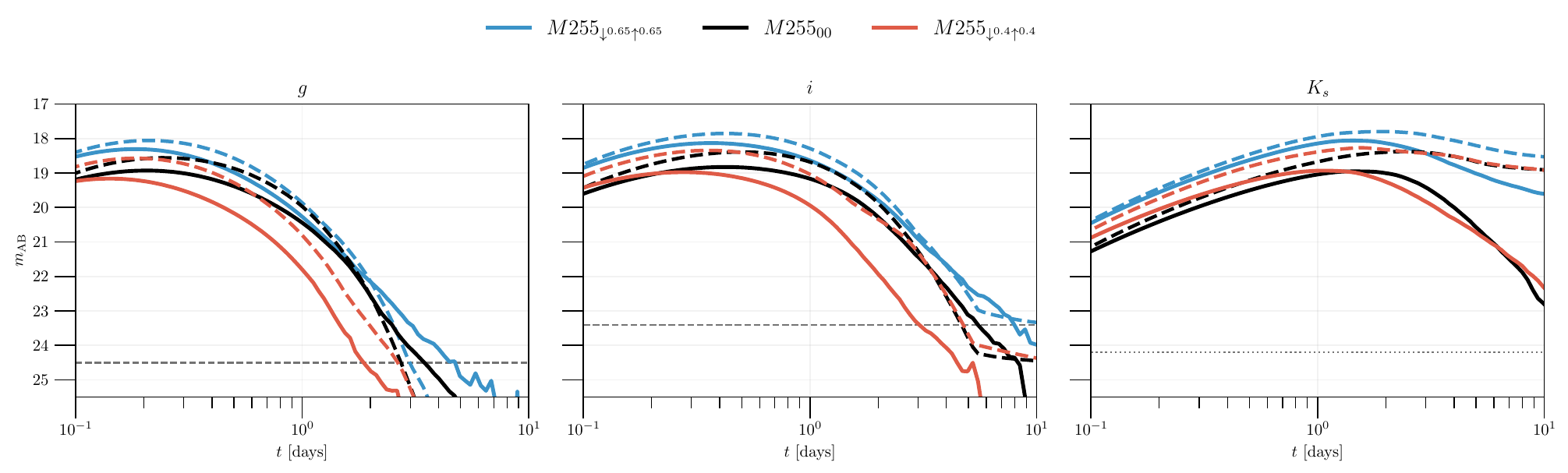}
    \caption{
    Apparent $g$, $i$ and $K_s$ kilonova light curves at $40\,\mathrm{Mpc}$ for the $M_\mathrm{tot}=2.55\,M_\odot$ models with $\chi_\mathrm{eff} = 0$. Solid lines show the simulation based ejecta, while dashed lines show the hybrid model with parametric disc outflows. The horizontal lines mark the adopted $5\sigma$ limiting magnitudes for Rubin $g$ and $i$ and VLT/HAWK-I in $K_s$.}
    \label{fig:kilonova_effspin}
\end{figure*}

The fixed $\chi_\mathrm{eff}=0$ colour differences also persist when the angular entropy and expansion timescale profiles are replaced by the mass weighted mean entropy and $\tau=1/v$. Replacing the simulation based secular ejecta with a parametric disc outflow reduces the peak brightness differences, but does not eliminate the fixed $\chi_\mathrm{eff}=0$ imprint, with peak brightness separations of up to ${\sim}0.6$ mag remaining, as shown in Fig.~\ref{fig:kilonova_effspin}. In this hybrid model, $M255_{00}$ and $M255_{\downarrow^{0.4} \uparrow^{0.4}}$ become nearly degenerate in peak brightness while retaining substantial peak associated colour differences.

When the same parametric disc outflow is applied to the fixed $\chi_\mathrm{eff}=0$ models using their respective disc masses, substantial peak associated colour differences remain between the irrotational and mixed models. The residual colour imprint persists despite the relatively small variation in the disc masses, indicating that the colour differences are not primarily driven by the disc mass variation. Differences in the dynamical ejecta beyond their total mass therefore appear to be an important source of the individual spin imprint in the kilonova colours at fixed $\chi_\mathrm{eff}=0$.

At fixed individual spin magnitudes, changing the spin orientation from anti-aligned to aligned changes the kilonova peak brightness by more than $1$ mag in all three bands for every viewing angle, and the corresponding single-spin comparison shows the same behaviour. The peak associated $i-K_s$ difference between the $M255_{\downarrow^{0.4} \downarrow^{0.4}}$ and $M255_{\uparrow^{0.4} \uparrow^{0.4}}$ retains the same sign over the full viewing angle range and reaches ${\sim}1.25$ mag, whereas the corresponding single-spin comparison reaches only ${\sim}0.15$ mag and does not retain the same sign across all viewing angles. The effect of increasing the spin magnitude also depends on the spin configuration and viewing angle. The brightness ordering reverses in the $g$ band for the aligned models and in $K_s$ for the anti-aligned models, while remaining unchanged in all three bands for the mixed models. The brighter kilonova emission of our anti-aligned models is qualitatively consistent with the trend reported by~\citet{East_spin_2019}, whereas~\citet{dietrich_spin_2017} found the opposite dependence, with aligned spin producing brighter EM counterparts. Earlier spinning BNS studies interpreted spin induced changes in the kilonova primarily through changes in the dynamical ejecta mass~\citep{dietrich_spin_2017, East_spin_2019}, while~\citet{East_spin_2019} also discussed the role of spin-induced changes in the disc properties for the expected blue kilonova. In our models, the dynamical ejecta mass alone does not uniquely determine the peak brightness, while substantial colour differences between the irrotational and mixed models persist when the same parametric disc outflow is applied using their respective disc masses, including at fixed $\chi_\mathrm{eff}$.

Neutrino absorption increases the model to model spread in peak brightness in the $i$ and $K_s$ bands at every viewing angle, by factors of ${\sim}1.4$-$2.4$ and ${\sim}1.8$-$1.9$, respectively. In the $g$ band, the spread increases for $\theta_\mathrm{obs}\geq30^\circ$ and reaches a factor of ${\sim}2.2$ toward equatorial viewing angles. Including neutrino absorption makes all five models brighter in all three bands at every viewing angle, with $\Delta M_g^\nu > \Delta M_i^\nu > \Delta M_{K_s}^\nu$ for every model and viewing angle, and shifts both the peak associated $g-i$ and $i-K_s$ colours blueward in all cases, by up to ${\sim}2.5$ mag. The magnitude and viewing angle dependence of this colour shift nevertheless vary substantially among the spin configurations. The $M255_{\uparrow^{0.4} \uparrow^{0.4}}$ model shows the strongest viewing angle dependence of the neutrino absorption effect, with the $g$ band brightening varying by ${\sim}1.4$ mag across the viewing angle range.

For $M_\mathrm{tot} = 3.05\,M_\odot$, the $q=1$ models do not show the same brightness ordering in the three bands. The $M305_{\downarrow^{0.4} \downarrow^{0.4}}$ model is the brightest in $g$ at every viewing angle, while $M305_{\uparrow^{0.67} \uparrow^{0.67}}$ is the brightest in $K_s$. In the $i$ band, the ordering changes with viewing angle. Among the aligned models, $M305_{\uparrow^{0.67} \uparrow^{0.67}}$ is brighter than $M305_{\uparrow^{0.4} \uparrow^{0.4}}$ in all three bands at every viewing angle.

The dependence on the individual spin configuration remains strong in the prompt collapse $M410$ models. The mixed  model is brighter than the aligned model in all three bands at every viewing angle, with polar differences reaching ${\sim}1.5$ mag. In $i$ and $K_s$, the viewing angle trend reverses between the two models, with the mixed model brightest toward the pole and the aligned model toward the equator. In $K_s$, $\Delta_\theta M_\mathrm{pk}$ is ${\sim}0.2$ mag for the mixed model and ${\sim}1.2$ mag for the aligned model. Previous work on irrotational binaries found that kilonovae from prompt collapse models are more strongly influenced by the dynamical ejecta and linked their properties to remnant lifetime and disc outflows~\citep{radice_massejection_2018}. Our results show that the individual spin configuration can produce substantial differences in the kilonova emission even among prompt collapse models.

\section{Conclusions}\label{conclusion}

In this study, building on the spinning binary neutron star merger simulations presented in~\cite{effofspin}, we focus on how the neutron star spins affect the ejecta and the resulting electromagnetic counterpart. To investigate these effects in more detail, we decompose the ejecta into two main components: $i)$ the dynamical ejecta, whose mass rapidly saturates, and $ii)$ the secular ejecta, which develops over longer timescales. We further separate the dynamical ejecta into tidal and shock-heated components. This decomposition allows us to better identify the origin of the $r$-process nucleosynthesis and to interpret the associated kilonova emission.

At LR, our simulations show that, for the $M_{\mathrm{tot}} = 2.55\,M_\odot$ models, the irrotational and anti-aligned configurations are shock-dominated, whereas the aligned, single-spin aligned, and mixed-spin configurations are dominated by tidal mass ejection. The single-spin anti-aligned model, in contrast, shows nearly equal contributions from the shock-heated and tidal components. This suggests that anti-aligned spins reduce the total angular momentum and favour stronger shock heating, whereas aligned spins favour tidal disruption. The relative contributions of the tidal and shock-heated ejecta in the anti-aligned and single-spin anti-aligned models are, however, sensitive to resolution and differ between LR and HR. Nevertheless, at fixed $\chi_\mathrm{eff}=0$ and at both resolutions, the irrotational model remains shock-dominated, whereas the mixed-spin model remains tidal-dominated. Physically, this shows that spin affects not only the total dynamical ejecta mass, but also the dominant mass-ejection mechanism. Most importantly, the dominant ejection channel is not determined by $\chi_{\mathrm{eff}}$ alone, but depends on the individual spin configuration.

For the $M_{\mathrm{tot}}=3.05\,M_\odot$ models, we find that the mass ratio can change the dynamical ejecta mass at fixed $\chi_\mathrm{eff}$, while the dominant mass-ejection channel changes only at $\chi_{\mathrm{eff}}=0$. More importantly, for the $M_{\mathrm{tot}}=4.10\,M_\odot$ models, although all configurations undergo prompt collapse to a black hole, the ejecta remain strongly dependent on the spin configuration. At fixed $\chi_{\mathrm{eff}}=0$, the mixed-spin model produces substantially more dynamical ejecta than the irrotational model.

It is also important to note that, while previous studies generally find only small amounts of fast-moving ejecta with $v>0.6\,c$ in irrotational BNS mergers~\citep{radice_massejection_2018, 2025MNRAS.538..907R} and in both non-spinning and spinning BHNS mergers~\citep{matur_bhns_spin_2025}, some of our anti-aligned spinning models produce fast-ejecta masses approaching ${\sim}10^{-3}\,M_\odot$. In particular, increasing the magnitude of the anti-aligned spin enhances the high-velocity ejecta with $v>0.3\,c$. For the high-spin anti-aligned $M_{\mathrm{tot}}=2.55\,M_\odot$ model, the fast ejecta reach a mean velocity of ${\sim}0.72\,c$.

We also present the secular ejecta for all models. For the $M_{\mathrm{tot}}=2.55\,M_\odot$ models, as in the dynamical ejecta, the individual spin configuration strongly affects the secular ejecta mass and properties. The high-spin aligned model produces the largest secular ejecta mass at both resolutions, reaching ${\sim}(2.5$--$2.9)\times10^{-2}\,M_\odot$. This dependence remains strong in the $M_{\mathrm{tot}}=4.10\,M_\odot$ prompt-collapse models: the aligned model produces nearly $40$ times more secular ejecta than the mixed-spin model, while the secular ejecta mass spans nearly four orders of magnitude across the models. These results indicate that the imprint of the individual spin configuration persists beyond the dynamical merger phase and strongly affects the subsequent secular outflows, even for systems with the same prompt-collapse fate.

The quantitative results should be interpreted taking into account the numerical uncertainties discussed in Sec.~\ref{setup} and Sec.~\ref{sec:resolution_dependence}. Some secular ejecta masses remain lower limits. The secular ejecta are more resolution sensitive than the dynamical ejecta. The comparisons across all models are based on the LR ejecta, while  the $M_\mathrm{tot}=2.55\,M_\odot$ models are also analysed at HR to assess resolution dependence. Nevertheless, the individual spin imprint persists in the dynamical ejecta, heavy $r$-process yields, and kilonova observables at both resolutions.

We also calculate the $r$-process nucleosynthesis for all ejecta channels. At fixed $\chi_{\mathrm{eff}}=0$, the heavy $r$-process mass fraction and absolute yield vary substantially between models because of their different individual spin configurations. In particular, the absolute yield of nuclei with $A\geq140$ differs by more than two orders of magnitude, while the lanthanide-to-light $r$-process mass ratio increases from ${\sim}2$ in the irrotational model to ${\sim}70$ in the high-spin mixed model. These differences become even larger in the $M_{\mathrm{tot}}=4.10\,M_\odot$ prompt-collapse models, where the absolute heavy $r$-process yield differs by more than four orders of magnitude at fixed $\chi_{\mathrm{eff}}=0$.

Examining the individual ejecta channels shows that the heavy $r$-process abundance pattern of the tidal ejecta remains relatively stable across spin configurations, whereas the shock-heated component is substantially more sensitive to spin. The final abundance pattern is therefore determined both by the relative contributions of the different ejecta channels and by spin-dependent composition changes within the individual channels.

In the kilonova calculations, we again find a strong dependence on the individual spin configuration. At fixed $\chi_\mathrm{eff}=0$, the peak brightness differences reach ${\sim}0.9$ mag, while the same-epoch colour differences reach ${\sim}1.5$ mag. At $40\,\mathrm{Mpc}$, all fixed $\chi_\mathrm{eff}=0$ model pairs have jointly observable epochs for every viewing angle considered, suggesting that the individual spin imprint could potentially be observable. Replacing the simulation-based secular ejecta with a parametric disc outflow reduces these differences but does not eliminate them, with peak brightness separations of up to ${\sim}0.6$ mag remaining. This strengthens the conclusion that differences in the dynamical ejecta make an important contribution to preserving the individual spin imprint in the kilonova emission.

For the $M_\mathrm{tot} = 2.55\,M_\odot$ models evolved at both resolutions, the dynamical ejecta mass spans factors of ${\sim}45$ at LR and ${\sim}30$ at HR, while the heavy $r$-process yield spans more than two orders of magnitude at both resolutions. The individual-spin dependence of the kilonova also persists, with peak magnitudes changing by at most ${\sim}0.5$ mag between LR and HR.

As expected, neutrino absorption increases the electron fraction of the ejecta, making the material less neutron-rich and reducing the relative production of heavy $r$-process nuclei. In the kilonova calculations, neutrino absorption systematically brightens the emission and shifts the peak-associated colours blueward.

Overall, our results show that the ejecta, $r$-process nucleosynthesis, and kilonova emission from spinning BNS mergers depend on the individual spins, $\chi_1$ and $\chi_2$, differently from the gravitational wave signal, for which the leading spin dependence is captured by $\chi_{\mathrm{eff}}$. The individual spin configuration affects not only the amount of mass ejection, but also the dominant ejection channel, the nucleosynthetic outcome, and the resulting electromagnetic counterpart, with these differences persisting even in prompt-collapse systems and in kilonova calculations. These results show that EM observations can break the degeneracy between individual spin configurations with the same $\chi_\mathrm{eff}$.

Future work should employ GPU-accelerated numerical relativity codes to perform higher-resolution simulations covering longer pre- and post-merger phases, enabling more accurate modelling of both the gravitational wave signal and the electromagnetic counterpart. In particular, magnetic-field effects are expected to become increasingly important over longer post-merger timescales. Investigating the interplay between spin and strong neutron-star magnetic fields will therefore be essential for improving predictions of the electromagnetic emission.

\section*{Acknowledgements}

This work used the DiRAC Memory Intensive service (Cosma8) at Durham University via RAC18 allocation (project ID dp437), managed by the Institute for Computational Cosmology on behalf of the STFC DiRAC HPC Facility (www.dirac.ac.uk). The DiRAC service at Durham was funded by BEIS, UKRI and STFC capital funding, Durham University and STFC operations grants. DiRAC is part of the UKRI Digital Research Infrastructure.
D.R. acknowledges support from NASA through Grant No. 80NSSC25K7213, from the Department of Energy, Office of Science, Division of Nuclear Physics, under Award Number DE-SC0024388, and from the National Science Foundation under Grants No. PHY‐2407681, PHY-2512802, PHY‐2608045, and PHY‐2621752. 
    
\section*{Data Availability}

The ejecta data presented in this work are publicly available on Zenodo~\citep{imprint_zenodo}.

\bibliographystyle{mnras}
\bibliography{references} 

\bsp
\label{lastpage}
\end{document}